\documentclass[aps,prb,twocolumn,groupedaddress,showpacs]{revtex4-1}
\usepackage{mathbbol,amsmath,amsfonts,amssymb,bm,color,dsfont}
\usepackage{graphicx,psfrag,array}

\def\tr{{{\sf Tr}}}

\providecommand{\ignore}[1]{}
\providecommand{\aucmnt}[1]{#1}
\renewcommand{\aucmnt}[1]{}

\newcommand{\ket}[1]{| #1 \rangle}

\newcommand{\Comment}[1]{}

\newcommand{\Eq}[1]{Eq.~(\ref{#1})}

\begin{document}

\title{Repulsive Interactions in Quantum Hall Systems as a Pairing Problem}

\author{G. Ortiz$^1$}
\email[Electronic address: ]{ortizg@indiana.edu}
\author{Z. Nussinov$^2$}
\author{J. Dukelsky$^3$}
\author{A. Seidel$^2$}
\affiliation{$^1$Department of Physics, Indiana University, Bloomington,
IN 47405, USA}
\affiliation{$^2$Department of Physics, Washington University, St.
Louis, MO 63160, USA}
\affiliation{$^3$Instituto de Estructura de la Materia, C.S.I.C., Serrano 123, 
E-28006 Madrid, Spain}

\date{\today}
\begin{abstract}

A subtle relation between Quantum Hall physics and the phenomenon  of
pairing is unveiled.  By use of second quantization, we establish a
connection between (i) a broad class of rotationally symmetric two-body
interactions within the lowest Landau level and (ii) integrable
hyperbolic Richardson-Gaudin type Hamiltonians that arise in
$(p_{x}+ip_{y})$ superconductivity. Specifically, we show that general
Haldane pseudopotentials (and their sums) can be expressed as a sum of 
repulsive non-commuting $(p_{x}+ip_{y})$-type pairing Hamiltonians.  The
determination of the  spectrum and  individual null spaces of each of
these non-commuting Richardson-Gaudin type Hamiltonians is non-trivial
yet is Bethe Ansatz solvable.  For the Laughlin sequence,  it is
observed that this problem is frustration  free and zero energy ground
states lie in the  common null space of all
of these non-commuting Hamiltonians. This property allows for the use of
a new truncated basis  of pairing configurations in which to express
Laughlin states at general filling factors.  We prove separability of
arbitrary Haldane pseudopotentials, providing   explicit expressions for
their second quantized forms, and further show by  explicit construction
how to exploit the topological equivalence between  different geometries
(disk, cylinder, and sphere) sharing the same topological genus number,
in the second quantized  formalism, through similarity transformations.
As an application of the  second quantized approach, we establish a
``squeezing principle'' that  applies to the zero modes of a general
class of Hamiltonians, which includes  but is not limited to Haldane
pseudopotentials. We also show how one may establish (bounds on)
``incompressible  filling factors'' for those Hamiltonians. By invoking
properties of symmetric polynomials, we provide explicit second
quantized quasi-hole generators;  the generators that we find directly
relate to bosonic chiral edge modes and further make aspects of
dimensional reduction in the Quantum Hall systems precise.


\end{abstract}
\maketitle

\section{Introduction}

The study of strongly correlated phases of matter is a supremely
challenging task, and yet has let to some of the most notable triumphs
in condensed matter physics. A major route to success along these lines
has been the identification of special points within a given phase of
interest, where the Hamiltonian becomes somewhat tractable and leads to
sufficiently simple ground and excited state wave functions, either
through exact or approximate treatment, that nonetheless capture the
essential universal features of the phase as a whole. The fractional
Quantum Hall (FQH) regime harbors a wealth of examples of the latter
kind. Indeed, in the presence of a strong magnetic field, powerful
principles are known constraining the construction of successful trial
wave functions\cite{laughlin, jain, Moore} as well as model
Hamiltonians.\cite{Haldane}  Moreover, for many (though certainly not
all) known trial wave functions describing various phases (or critical
points\cite{gaffnian}) in the FQH regime,  a local model Hamiltonian can
be identified for which the wave function in question is an exact ground
state. The identification of such {\it parent} Hamiltonians is usually
greatly aided by special analytic properties of the underlying wave
function.  The primary example where this scheme of attack has been
successful is the $\nu=1/3$ Laughlin state, whose parent Hamiltonian is
the $V_1$ Haldane pseudopotential.\cite{Haldane, Trugman}  For fermions,
the latter can be regarded as the simplest local two-body interaction
within the lowest Landau level (LLL).

Despite these powerful applications, the dependence on analytic wave
function properties in the construction of Quantum Hall (QH) parent
Hamiltonians also leads to severe inherent limitations. For one, many QH
phases of fundamental interest, such as those described by
hierarchy\cite{Haldane, halperin} or Jain\cite{jain} composite fermion
states, do not have known parent Hamiltonians. Moreover, as was recently
argued by Haldane,\cite{Haldane2011} all information about the
topological order of the ground states is encoded in the guiding center
degrees of freedom only, whereas analytic  properties of the wave
function are due to the interplay of  the latter with the particles'
dynamical momenta, which determine the structure of a given Landau
level. This structure is arguable not essential for the topological
quantum order of the QH  fluid. Indeed, as we will review below, in a
strong uniform magnetic field one may formulate the Hamiltonian dynamics
of the electrons in a second-quantized ``guiding-center only'' 
language, which is stripped of the dynamical momenta entirely. As is well
appreciated, QH physics is intimately tied to  dimensional reduction
which is similarly manifest in many other systems exhibiting topological
orders. \cite{NO} In the associated ``guiding-center only'' second
quantized Hamiltonian (wherein the two spatial dimensions of the original
QH problem in first quantization are replaced by a one-dimensional (1D)
fermionic lattice of the angular momentum orbitals) this dimensional
reduction becomes explicit and leads to a class of 1D lattice models
which may be of interest in a more general context outside Landau-level
physics (see Fig. \ref{DimRed.fig}). Specifically, this has been
proposed  recently for flat band solids with and without Chern
numbers.\cite{Qi, Scarola} 

For all the above reasons, it is desirable to understand  the existing
parent Hamiltonians of FQH model wave functions in the context
of 1D lattice Hamiltonians, \cite{1Dparent,seidel,berg,Nakamura}  and
most importantly the  QH projected Hamiltonian in second
quantized  form.  Due to technical difficulties that we will elaborate on
below (see also Ref. \onlinecite{zhou12}), there currently seems to be
very limited understanding of how the (quasi)-solvability of FQH
parent Hamiltonians follows from its defining operator algebra  in the
1D lattice or ``guiding center'' picture.

In this work, we aim to improve on the above situation. We expose a
connection between the operator algebra defining the parent Hamiltonian
of Laughlin QH states on the one hand, and another paradigmatic state of
strongly correlated matter on the other, the superconductor. More
precisely, we expose that a {\it generic two-body interaction  can be
written as a direct sum of hyperbolic Richardson-Gaudin (RG)
Hamiltonians} in the strongly coupled repulsive regime. \cite{NuclPhysB}
Hyperbolic RG models represent a general class of exactly solvable  {\it
pairing} Hamiltonians  which includes the $p_x+ip_y$ superconductor  as
a particular  instance. \cite{Romb}   We use the fact that  the latter
(generally non-commuting) RG  Hamiltonians are exactly solvable (by
Bethe-Ansatz \cite{NuclPhysB, Romb}) to characterize the individual
non-trivial null spaces of these RG operators at a given filling $\nu$.
The common null space of the latter is the corresponding  Laughlin
state. Thus, when expressed as a sum of RG Hamiltonians, the
``frustration free'' character of the  system is underscored.  A
Hamiltonian is termed ``frustration free'' (or ``quantum satisfiable'')
whenever all of its null states are also null states of each of the
individual terms (in our case, RG Hamiltonians) that form it. That is,
the ground states of any individual RG Hamiltonian are globally
consistent with the ground state null space of the full Hamiltonian.
Frustration free Hamiltonians have recently been under intense study at
the interface of condensed matter and quantum information theory.
\cite{frustration_free}  We remark that while most frustration free
Hamiltonians studied in the literature are sums of similar {\it local}
Hamiltonians which merely operate on different sites (e.g., local
Hamiltonians related to one another by lattice translations), those in
the QH problem are richer; the RG Hamiltonians which form the full 
QH system that we consider are not strictly finite ranged. Thus, unlike 
in simple lattice models, the study of their eigenstates and eigenvalues 
is already a rich and non-trivial problem. We further explicitly note that
the aforementioned {\it strongly repulsive} $(p_x+ip_y)$-type RG
Hamiltonians that we will study, which share conventional Laughlin states
as their common ground states are notably very different from the far
more exotic Pfaffian type states for other fillings  and viable
insightful links to superconductivity therein. \cite{Moore,Read}  Our
hope is that by understanding the common null space of {\it all} 
hyperbolic RG Hamiltonians we will be able to shed light on  QH fluids
with filling fractions $\nu$ other than those with Laughlin type ground
states. 

Aside from considerations regarding effective field theories, the QH
problem seems to  find its most common representation in a first
quantized language of known (or guessed) ground states where properties
of holomorphic functions can be elegantly employed.  The second
quantized formulation, on the other hand,  sheds light on the algebraic
structure of QH system and does not explicitly rely on prior knowledge
of the form of the ground states. It further allows for the study of
excitations  above the ground states.  
As is
well known, only the genus number  sets the system's degeneracy
\cite{wen} (a feature which largely first triggered interest in
topological orders). 
We find it useful to recover this statement within second quantized formalism,
by constructing a similarity transformation that relates frustration free
eigenstates in disk, cylinder and sphere geometries.
Our analysis is valid for general frustration free Hamiltonians 
at arbitrary filling fractions and, as noted above, illustrates that general
interactions within the LLL can be expressed as a sum of RG type
Hamiltonians. We explicitly provide expressions for the second quantized
Haldane pseudopotentials in disparate geometries and find that
individual pseudopotentials  have a simple separable structure.  Lastly,
the quasi-hole operators that we find within second quantization have a
canny similarity to operators in 1D bosonized systems and further
suggest rigorous links  to dimensional reductions and earlier notions
regarding chiral edge states.

This work is organized as follows: In Section \ref{2nd_quant}, we setup
generic two-body QH Hamiltonians in  second quantized language. We start
(Section \ref{prelim}) by discussing general aspects of interactions
within the LLL. We then turn in Section \ref{V_1} to the lowest order
Haldane pseudopotential  (the Trugman-Kivelson model \cite{Trugman}) and
show that this Hamiltonian obtains a simple separable structure in second
quantization.  In Section \ref{V_n}, we provide the second quantized
form of {\it all two-body} Haldane pseudopotentials in disparate
geometries. Moreover,  we show that all, i.e., arbitrary order, Haldane
pseudopotentials are  {\it separable}, a key result for what follows. 
We then illustrate (Section \ref{qha}) that general QH Hamiltonians are
described by  an affine Lie algebra without a central extension.
Equation (\ref{HLLL}) will prove to be of immense use in our analysis in
later sections and allow to illustrate how the LLL Hamiltonian may be
written as a sum of individual RG type Hamiltonians, which include the 
strongly-coupled limit of the  $(p_{x}+ip_{y}$) superconductor.  In
Section \ref{equivalence}, we illustrate how similarity transformations
may exactly map QH systems on different surfaces (e.g., cylinder and sphere) when all of these
surfaces share the same genus numbers.  

In Section \ref{strong}, building on the decomposition of Eq.
(\ref{HLLL}), we discover a profound connection between a general (i) QH
system on the righthand side of Eq. (\ref{HLLL}) and (ii) repulsive
$(p_{x}+ip_{y})$-type RG Hamiltonians (which as we show correspond to
fixed values of $j$ on the righthand side of Eq. (\ref{HLLL})) and use
this relation to construct our framework for investigating the QH
problem. The role of pairing within the RG approach becomes apparent. 
Specifically, in Section \ref{exact} we demonstrate, via an exact
mapping, the above connection to the RG problem as it appears for each
individual value of the angular momentum $j$ and $m$. In Section
\ref{Hilbert}, we analyze the Hilbert space dimension associated with
the RG basis by building on links to generating functions and  a problem
of constrained non-interacting spinless fermions. We then proceed
(Section \ref{eigenspectrum}) to provide a {\it new} Bethe Ansatz
solution to the non-trivial spectral problem associated with the RG
Hamiltonian $H_{{\sf G}j;m}$, dubbed QH-RG, appearing for a fixed value
of $j$ and $m$. We discover two  classes of solutions, one associated to
a highly degenerate zero energy  (null) subspace and another with a well
defined sign of the eigenvalues.  The RG Hamiltonians generally do not
commute with one another.  We then turn to symmetry properties of this
new RG problem in Section \ref{sym_prop}. 

Equipped with an understanding of the RG problem, we next turn (Section
\ref{full_QH}) to the full QH problem which is a sum over such QH-RG
Hamiltonians. In Section \ref{nullspace}, we discuss general properties
of the common null space of the individual QH-RG Hamiltonians and
highlight the frustration free character of the QH problem when viewed 
through the prism of decomposition into non-commuting QH-RG Hamiltonians
(each of which has its own null space). Next (Section \ref{inward}), we
explicitly make use of second quantization and prove results concerning
the form of QH ground states from that perspective. In particular, for the 
zero modes of a general class of Hamiltonians, we rigorously establish 
constructs involving ``squeezing" and generalized Pauli-principles. 
  In Section \ref{RG0},
we highlight the viable use of the RG basis in writing down ground
states of the QH system. To make the discussion very tangible, we
discuss a simple explicit example,  that of $N=2$ particles within the
$\nu =1/3$ Laughlin state. In Section \ref{slatersec}, we review
rudiments of the currently widely used Slater decomposition basis. In
this basis, the role of pairing is highlighted and we review how
admissible states in the Slater determinant decomposition are related to
those obtained by ``squeezed-state'' considerations. 

In Section \ref{2ndhole}, we return to more general aspects and
illustrate how the power sum generating system of symmetric polynomials
enables  us to exactly write down the second quantized form of
quasi-hole creation operators. As with nearly all of the results that we
report in our work, this second quantized form that we obtain is exact
for a general number of particles $N$ and readily suggests links to
bosonized forms associated with chiral edge states. We conclude, in
Section \ref{conclusions}, with a brief synopsis of our results.
Additional technical details  concerning the derivation of the
coefficients appearing in the second quantized form  of the
pseudopotential (Section \ref{2nd_quant}) in disk and sphere geometries
are relegated to Appendix \ref{appA}. In Appendix \ref{appB} we analyze 
the set of Slater determinants admissible in the expansion of Laughlin states, 
establishing an equivalence between Young tableaux and squeezing expansions. 
 
\section{Quantum Hall Hamiltonians in Second Quantization}
\label{2nd_quant}

FQH fluids are archetypical strongly interacting systems that exhibit
topological quantum order.  At general filling fractions, their analysis
has proven to be extremely rich. In the traditional approach, assumed
knowledge of the ground states motivates the construction of parent
Hamiltonians.
In this article, we deviate from this path. We explicitly
construct  the {\it second quantized} form of a general LLL QH
Hamiltonian in various geometries (disk, sphere, cylinder, and torus)
and, for the frustration free case that is often of interest, 
study properties of    
the ground states and excitations
about them rather generally. Towards this end, we study the Haldane
pseudopotentials of various orders, show that (quite universally) they
obtain a separable multiplicative form, and explicitly illustrate that
genus number preserving deformations that do not alter the system
topology can be exactly implemented via similarity transformations.
Perhaps most pertinent to future sections is the reduction of the
generic LLL Hamiltonian to the representation provided in Eq.
(\ref{HLLL}) with the algebra defined by the relations of Eq.
(\ref{Tj}). This latter result will prove crucial in our analysis and
reduction of the general QH problem to that as a sum of non-commuting RG
$(p_{x}+ip_{y})$-type Hamiltonians. 

\subsection{One-dimensional Hamiltonians in the orbital basis}
\label{prelim}

A QH system consists of $N $ electrons moving on a 2D surface in the
presence of a strong magnetic field $\mathbf{B}$ perpendicular to that
surface. Laughlin states, which describe  incompressible quantum fluids,
capture essential correlations for certain filling fractions $\nu$
[which, for the disk, cylinder, and sphere, we define as $\nu =
(N-1)/(L-1)$, while $\nu= N/L$ for the torus, where $L$  is the number
of occupied orbitals in the Laughlin state, see Tables \ref{table2} and
\ref{table:nonlin}]. 
\begin{table}[thb]
\begin{tabular}{ c|c|c|c|c}
\hline\hline
geometry  & disk  & cylinder & sphere  & torus \\ \hline &&&& \\ $\nu$&
$\frac{N -1}{L-1}$  & $\frac{N -1}{L-1}$  & $\frac{N -1}{L-1}$  &
$\frac{N }{L}$   \\
\end{tabular}
\caption{Filling fraction $\nu$ for various geometries. }
\label{table2}
\end{table}

\begin{figure}[htb]
\includegraphics[width=0.99\columnwidth]{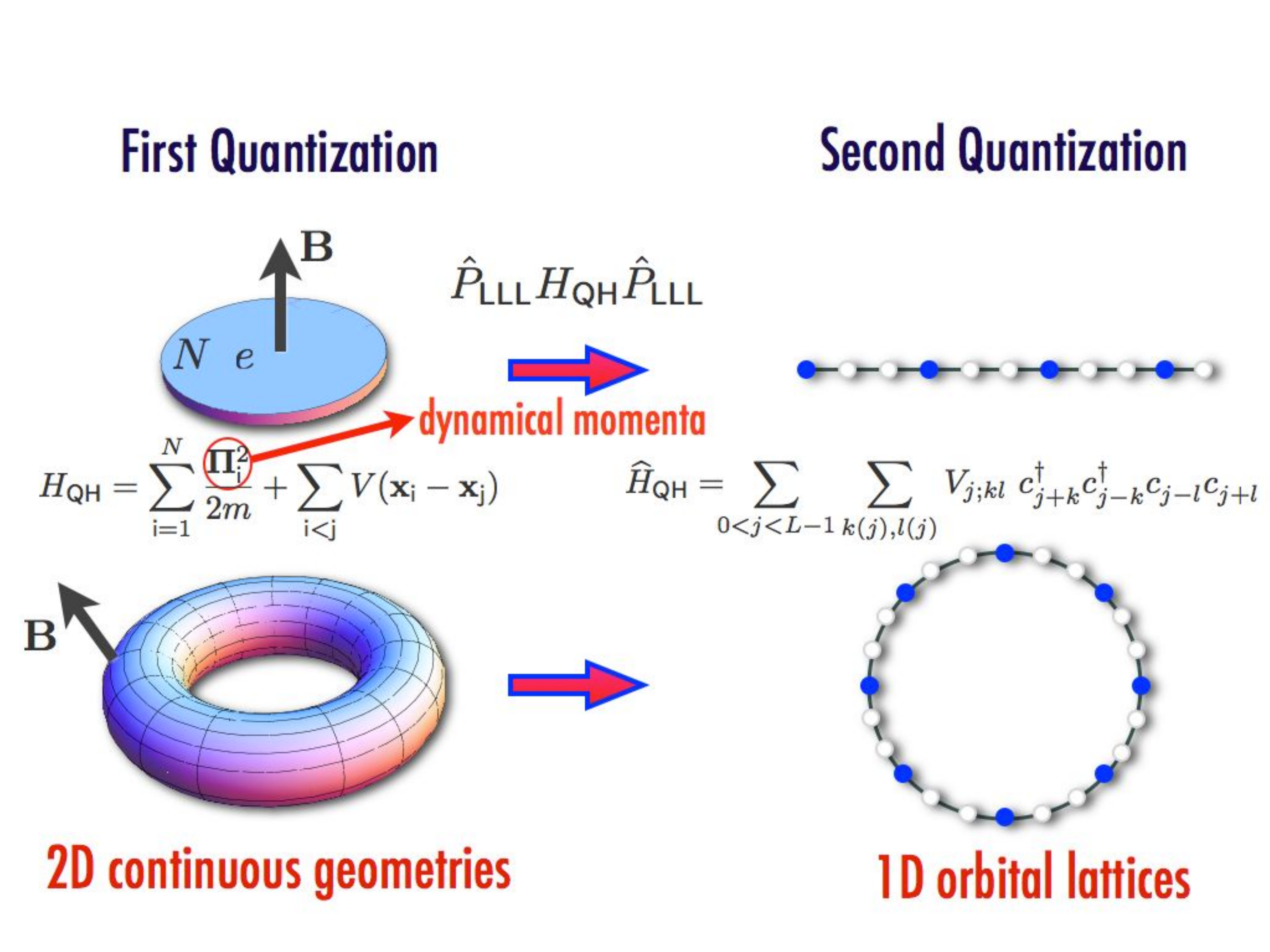}
 \caption{Pictorial representation of the effective dimensional reduction 
realized when going from the first to the second quantization 
representation of the QH Hamiltonian. }
 \label{DimRed.fig}
\end{figure}

The QH  Hamiltonian is given by $H_{\sf QH}=H_K+H_{\sf int}$, where
$H_K$ is the kinetic energy and only depends on the particles' dynamical
momenta, defining the degenerate Landau level structure. This degeneracy
is attributed to the particles' guiding center coordinates,  and at
non-integer filling fractions is only lifted by the interaction, which
we take to be of the following two-body form 
\begin{eqnarray}
\hspace*{-0.7cm}
H_{\sf int}&=&\! \frac{1}{2}\!  \int \!  d^2{\bf x}\, d^2{\bf x}'
\Psi^\dagger(\mathbf{x})\Psi^\dagger(\mathbf{x}') V({\bf x}-{\bf x}')
\Psi(\mathbf{x}')\Psi(\mathbf{x}) ,
\end{eqnarray}
with an interaction energy corresponding to a repulsive two-body
potential  $V({\bf x}-{\bf x}')$. The  field operator
$\Psi^\dagger(\mathbf{x}) = \sum_{r \in \mathbb{Z}} \phi_r^*(\mathbf{x})
c^\dagger_{r}$ is written in terms of fermionic operators
$c^\dagger_{r}$, creating fully polarized electrons in orbitals
$\phi_r(\mathbf{x})=\phi_r(z)$,  with orbital index $r$, where $\mathbf{x}=(x,y)$ and $z=x+i y$.
If $\mathbf{B}$ is strong enough, then to a reasonable approximation, we
may project, \cite{Haldane} onto the LLL (or any other Landau level)
consisting of $L$ orbitals.  Much of the physics of the QH effect can be
understood by such restricted dynamics.  With $\hat{P}_{\sf LLL}$
representing  the orthogonal projector onto the LLL, the kinetic energy
gets quenched and the relevant low-energy physics, in the presence of
rotational/translational symmetries, is described by
\begin{eqnarray}
\hspace*{-0.7cm}
\widehat{H}_{\sf QH}&=&\hat{P}_{\sf LLL}H_{\sf int}\hat{P}_{\sf LLL}
\nonumber \\
&=&
\sum_{0< j<L-1}\; \sum_{k(j),l(j)} V_{j;kl} \ c^\dagger_{j+k}
c^\dagger_{j-k} c^{\;}_{j-l} c^{\;}_{j+l}\;,\label{Hgen}
\end{eqnarray}
which describes an effective ``1D lattice system''
where
\begin{eqnarray}
\sum_{k(j)}=\sum_{0<k \leq \min(j,L-1-j)} ,
\label{ksum0}
\end{eqnarray}
and similarly for the sum over $l$. 
Here, the orbital indices  associated with this 1D lattice structure
refer to different states of the particle's guiding center (see Fig.
\ref{DimRed.fig}).  

In the sums defining the Hamiltonian  \eqref{Hgen} we leave implicit the
constraint that  orbital indices $j\pm k$ and $j\pm l$ are integer. 
This implies that these sums go over both triples  $(j,k,l)$ with all 
entries integer {\em and} triples $(j,k,l)$ with all entries half-odd
integer. With $j$ being restricted to the interval $[0,L-1]$, $j$ thus
takes on the $2L-3$ consecutive  values
\begin{eqnarray} \hspace*{-0.5cm}
j_{\sf min}=\frac{1}{2},1,\frac{3}{2},2,\dotsc, j_{\sf m}=\frac{L-1}{2},
\dotsc, j_{\sf max}=L-\frac{3}{2}.
\label{jvalues}
\end{eqnarray}
The sum over $k$  in Eq. \eqref{ksum0} starts at $k_{\sf min}=\frac{1}{2}$($1$) if
$j$ is half-odd integer(integer), ends at $k_{\sf max}= \min(j,L-1-j)$,
and involves 
\begin{eqnarray}
{\cal C}(j)=\min([j+\frac{1}{2}],[L-\frac{1}{2}-j])
\label{numbkl}
\end{eqnarray}
terms, with $[k]$ representing the integer part of $k$.   The ``middle''
allowed angular momentum value is  $j_{\sf m}$.  

\begin{table*}[hbt]
\begin{tabular}{ p{1.5cm}  m{2.1cm}  m{1.0cm}  m{6.1cm} m{6.2cm}}
\hline\hline
geometry  & $L$~ (Laughlin)  &  $N_\Phi$ & $\eta_{k}(j,1)$  &
{$\phi_r(z)$} \\ [0.5ex] 
\hline\\
disk & ${\sf q}N -{\sf q}+1$  & $L$ & $k\,2^{-j+1}\sqrt{\frac{1}{ j}{2j
\choose j+k}}$ & $\frac{1}{\sqrt{2\pi 2^r r!}} \, z^r
e^{-\frac{1}{4}|z|^2}$\\[0.5ex]
cylinder & ${\sf q}N -{\sf q}+1$ & $L$  & $2(8/\pi)^{1/4}\kappa^{3/2} \,k \,
e^{-\kappa^2 k^2}$ & $(4\pi^3)^{-1/4}\sqrt{\kappa}\, e^{-\frac{1}{2}
(x-r\kappa)^2+ir\kappa y }$ \\[0.5ex]
sphere& ${\sf q}N -{\sf q}+1$ & $L-1$  & $k \sqrt{\frac{2N_\Phi-2}{j(N_\Phi-j)}
{N_\Phi \choose j-k}{N_\Phi\choose j+k}/{2N_\Phi\choose 2j} }$
& $\sqrt{\frac{N_\Phi+1}{4\pi}{N_\Phi\choose
r}}[e^{-i\frac{\varphi}{2}}\sin(\frac{\theta}{2})]^r
[e^{i\frac{\varphi}{2}}\cos(\frac{\theta}{2})]^{N_{\Phi}-r}$\\[2ex]
torus & ${\sf q}N $ & $L$ & $\displaystyle 2(8/\pi)^{1/4}\kappa^{3/2} \,
\sum_{s\in\mathbb{Z}} (k+sL) \, e^{-\kappa^2 (k+ s L)^2}$ &
$\displaystyle \sum_{s \in\mathbb{Z}} \phi_{r +s L}^{\mbox{\scriptsize
cylinder}}$  \\[2ex]
\hline
\end{tabular}
\caption{\label{tab} Interactions  $\eta_{k}(j,1)$ for various
geometries. $L$ is the number of orbitals in incompressible QH systems
for filling fractions $\nu = 1/{\sf q}$, with ${\sf q}$ an odd integer. 
In the  disk geometry $\eta_{k}(j,1) \rightarrow  2\pi^{-\frac 14}
j^{-\frac 34} k\exp(-\frac{k^2}{2j})$ in the limit where $k\ll j$ and
not necessarily $k^2\ll j$.  The inverse radius of the cylinder or torus
(in the $y$-direction) is $\kappa=2\pi/L_y$. $N_{\Phi}$ is the number of
flux quanta threading the system. For  geometries without boundaries,
the relation between $N_{\Phi}$ and $L$ is unambiguous. We set $N_{\Phi}
= L$ for the disk and cylinder.
}
\label{table:nonlin}
\end{table*}

The structure of Eq. (\ref{Hgen}) preserves total angular momentum $J$.
The interaction Hamiltonian thus manifestly acts only on guiding center
variables, while leaving the dynamical momenta (related to the Landau
level index) invariant. The one dimensional sum in Eq. (\ref{Hgen})  is
intimately related to the dimensional reduction that the QH
system exhibits. Disparate systems that exhibit topological orders also
display dimensional reductions. \cite{NO,NOC-Holo} 

A broad class of rotationally invariant Hamiltonians in the LLL can be
written  as a sum of Haldane pseudopotentials ($\{H_{V_{m}}\}$) of order
$m=1,2, \ldots$,
\begin{eqnarray}
\label{sumQH}
\widehat{H}_{\sf QH} = \sum_{m>0} g_{m} H_{V_{m}},
\end{eqnarray} 
with general coefficients $g_{m}$.  In what follows, we first discuss
the  lowest order pseudopotential and then detail the algebraic
structure associated with it and all higher order pseudopotentials. 

\subsection{Lowest order Haldane pseudopotential or the Trugman-Kivelson
model}
\label{V_1}

It is a complex task  to analytically resolve the  spectral properties
of $\widehat{H}_{\sf QH}$. It is well-known that Laughlin states,
$\ket{\Psi_\nu}$, characterized by an odd integer $\sf q$  (${\sf q} \in
\{1,3,5,\cdots \}$), with $\nu=1/{\sf q}$ being the filling fraction in
the thermodynamic limit, are ground states of the {\it separable} lowest
order Haldane pseudopotential or Trugman-Kivelson Hamiltonian
\cite{Trugman,Haldane}
\begin{eqnarray}
\hspace*{-0.7cm}
H_{V_1}&=&  \sum_{0< j<L-1}\; \sum_{k(j),l(j)}
\eta_{k} \eta_{l} \ c^\dagger_{j+k}
c^\dagger_{j-k} c^{\;}_{j-l} c^{\;}_{j+l} \ ,
\label{V1QHE}
\end{eqnarray}
where the sums satisfy the same constraints as those in Eq.
\eqref{Hgen}. The summation is performed over $k,l$ in accord with the
convention of Eq. (\ref{ksum0}),  and the coefficients
$\eta_{k(l)}\equiv \eta_{k(l)}(j,1)$ depend on the geometry (see  Table
\ref{tab}).  

{}From the definition of the filling  fraction in QH systems
(see Tables \ref{table2} and \ref{table:nonlin}),
\begin{eqnarray}
\nu=\frac{N -1}{L-1}=\frac{N (N -1)}{2J_{\sf m}}=\frac{p}{q} \ ,
\label{fillingfrac}
\end{eqnarray}
where $J_{\sf m}=N  \, j_{\sf m}$ is the total angular momentum of the 
Laughlin state $\ket{\Psi_\nu}$ and $p,q$ are relatively prime integers.

We enforce a hard wall constraint on the disk and the cylinder that
limits the available Landau level orbitals to $L$ consecutive  orbitals.
For the compact sphere (and torus), the LLL is naturally finite
dimensional. With this, the $\nu=1/3$ Laughlin state  (with $N $ electrons
(see Table \ref{tab})) is the unique {\em zero energy} ground state of
the positive semi-definite Hamiltonian \eqref{V1QHE} for the disk,
cylinder, and spherical geometries. The  completely filled Landau level
$\nu=1$ fluid has a unique (ground)  state, which is  a simple Slater
determinant, but of positive energy.  For  $\nu< 1/3$, the Laughlin
state is still a zero energy ground state of \Eq{V1QHE}, but additional
pseudopotential terms are required to render this ground state unique.
\cite{Haldane} The most general radially symmetric  interaction
potential can be expressed as a sum of such pseudopotentials.

On the torus,  Eq. \eqref{V1QHE} must be modified to read
\begin{eqnarray}
\hspace*{-0.7cm}\label{HV1}
H_{V_1}&=& \sum_{0< j<L}\; \sum_{0<k,l<L/2}
\eta_{k} \eta_{l} \ c^\dagger_{j+k}
c^\dagger_{j-k} c^{\;}_{j-l} c^{\;}_{j+l} \,.
\label{V1QHEtor}
\end{eqnarray}
Again, all integer or all half-odd integer values $(j,k,l)$ are allowed
in the sum. Moreover, the operator $c_r$ is identified with $c_{r+L}$
due to periodic boundary conditions. These boundary conditions are also
respected by the symmetry of the  $\eta_{k}$ symbols on the torus,
$\eta_{k+L}=\eta_{k}$. Due to center of mass degeneracy, there are $\sf
q$ degenerate  $\nu=1/{\sf q}$ Laughlin states on the torus.  For ${\sf
q}=3$, these are the unique zero energy ground states of Eq.
\eqref{V1QHEtor}. \cite{Haldane_Rez}

\subsection{Separability of general pseudopotentials}
\label{V_n}

For simplicity, in later sections we may often have in mind the
coefficients $\eta_{k}$ that define the $V_1$ pseudopotential.
Nevertheless, as we will now illustrate, {\it all} 
higher ($m >1$) order
pseudopotentials  are of the same factorized form in terms of 
$\eta$ symbols
as  displayed in \eqref{V1QHEtor}.

We begin with the disk geometry, where in a first quantized language,
the pseudopotentials $V_m$ are defined as\cite{Haldane}
\begin{eqnarray}
  V_m=\sum_{{\sf i<j}} P_m({\sf ij})\,,
\end{eqnarray}
with $P_m({\sf ij})$ a projector acting on the pair of particles
$({\sf ij})$ and projecting onto the subspace  associated with relative
angular momentum $m$. In second quantization, using the basis of
single-particle angular momentum eigenstates, the most general form
$V_m$ can have is
\begin{eqnarray}\label{Vmsec}
   H_{V_m} &=&\sum_{0<j<L-1} H_{V_{m,j}}\\ \nonumber
   &=&  \sum_{0<j<L-1}\sum_{k(j),l(j)}  M^j_{kl;m}\; c^\dagger_{j+k}
   c^\dagger_{j-k}  c^{\;}_{j-l} c^{\;}_{j+l}\,,
\end{eqnarray}
where conservation of total angular momentum has been used, and
$H_{V_{m,j}}$ is the operator corresponding to fixed $j$ in the second
line. We note that the results of this section are valid for both fermions
and bosons, where for the latter case, we must replace Eq. \eqref{ksum0}
with the more symmetric definition, 
\begin{equation}
\sum_{k(j)}=\frac 12 \sum_{0\leq |k| \leq \min(j,L-1-j)}\,,\label{ksum}
\end{equation}
and similarly for $\sum_{l(j)}$.
Let us now consider the action of $H_{V_m}$ on states with two
particles. The operator $H_{V_{m,j}}$ apparently projects any pair state
with {\em total} angular momentum $2j$ onto a state with the same total
angular momentum, and, as we know from the definition, relative angular
momentum $m$. For two particles, however, the total and relative angular
momenta fully specify the state. Therefore, $H_{V_{m,j}}$ is the
projection onto the unique two-particle state specified by the quantum
numbers $m$, $j$. It follows from this that  $M^j_{kl;m}$ as a matrix in
$k,l$ for fixed $j$ must have rank $1$. It is further Hermitian and real
(by PT-symmetry). Its most general form is therefore given by
$M^j_{kl;m}=\eta_{k}(j,m)\eta_{l}(j,m)=\eta_k \eta_l$, where we leave
implicit  the $j$ and $m$ dependence of the $\eta$-symbols. Within the
two-particle subspace, the operator $H_{V_{m,j}}$ is thus the orthogonal
projection  onto the state
\begin{eqnarray}\label{jmstate}
\sum_{k(j)} \eta_{k} \, c^\dagger_{j+k}
 c^\dagger_{j-k}|0\rangle\,.
\end{eqnarray}
To characterize the spectrum and the eigenstates of $H_{V_{m,j}}$ 
within the general $N$-particle subspace will be a main focus of this
paper. The characterization of $H_{V_m}$ as a projection operator within
the two-particle subspace, however, also implies that the normalization
of the state \eqref{jmstate} must be unity independent of $j$:
\begin{eqnarray}\label{etanorm}
 \sum_{k(j)}(\eta_{k})^2=1\quad ({\sf for}\;\; 2j\geq m)\,.
\end{eqnarray}
The restriction is due to the fact that for $2j<m$, $\eta_{k}
\equiv 0$ as will  presently become apparent. An explicit formula for
$\eta_{k}$ can be obtained from Eq. \eqref{jmstate} and the fact
that the normalized first quantized two-particle wave function  of
relative angular momentum $m$ and total angular momentum $2j$ is
\begin{equation}\label{jmstate1st}
  \frac{2^{-2j}}{2\pi\sqrt{(2j-m)!
m!}}\left({z_1+z_2}\right)^{2j-m}(z_1-z_2)^m e^{-\frac 14 |z_1|^2-\frac
14 |z_2|^2}\,,
\end{equation}
so long as $2j\geq m$. (There exists no such state otherwise.)
Expressing this in second quantization, one obtains 
\begin{subequations}
 
\begin{equation}
\begin{split}\label{hyper}
\eta_{k}= &(-1)^{m+j-k}\sqrt{\frac{(j-k)!(j+k)!}{{2^{2j-1}(2j-m)!
m!}}}\;\times \\
& \hspace*{-0.8cm} {m \choose j-k} {_2} F_1 (-j+k,-2j+m,1-j+k+m,-1)
\end{split}
\end{equation}
where ${_2}F_1$ is a hypergeometric function. For later purposes, it is
important to note that the last equation is of  the following structure

\begin{equation}\label{etadisk}
 \eta_{k}=2^{-j+1/2}\sqrt{{2j \choose m}{2j \choose j+k}} \ p_{m,j}(k)\,,
\end{equation}
\end{subequations}
where $p_{m,j}(k)$ is a polynomial in $k$ of degree $m$ and parity
$(-1)^m$. We will prove this in Appendix \ref{appA} and also give a
recursive formula for the $p_{m,j}$. 

We note the orthogonality of the states \eqref{jmstate1st} for different
$j$, $m$. Working still at fixed $j$ and making the $m$-dependence of
the $\eta_k$ explicit for now, we have, on top of  Eq. \eqref{etanorm}:
\begin{eqnarray}\label{etaorth}
   \sum_{k(j)} \eta_{k}(j,m)\eta_{k}(j,m')=\delta_{m,m'}\;\;({\sf for}\;\;
2j\geq m,m')\,.
\end{eqnarray}
This observation will directly carry over to the sphere, but not to the
cylinder or torus.

For the sphere, the situation is very similar, except $P_m({\sf ij})$
must be defined as the projection of particles $\sf i$ and $\sf j$ onto
the two-particle  subspace of {\em total} angular momentum ${\sf
L}^2={\sf \ell}( {\sf \ell}+1)$ with ${\sf
\ell}=N_\Phi-m$.\cite{Haldane} The quantum number $j$ now corresponds to
the $z$-component of angular momentum, where $2j$ is the total ${\sf
L_z}$ of the pair. 
Again this uniquely specifies a two-particle state. The same argument as
given above for the disk then implies the separable form of $H_{V_m}$ in
second quantization. Moreover, noting that each individual particle in
the LLL transforms under the spin $N_\Phi/2$ representation of SU(2),
the coefficient $\eta_{k}$ defining the state \eqref{jmstate} for
given $j$ and $m$ is simply a Clebsch-Gordan coefficient, or, written as
a 3$j$-symbol,
\begin{subequations}
\begin{eqnarray}\label{etasphere0}
    &&\eta_{k}= (-1)^{N_\Phi-2j}\sqrt{4N_\Phi-4m+2} \times \nonumber
\\
   && \begin{pmatrix} N_\Phi/2 & N_\Phi/2 & N_\Phi-m \\
  N_\Phi/2-j+k  &  N_\Phi/2-j-k & 2j-N_\Phi 
 \end{pmatrix}.
\end{eqnarray}
Again, we note for later purposes the analog of Eq. \eqref{etadisk} for
the sphere:
\begin{equation}\label{etasphere}
 \eta_{k}=\sqrt{{N_\Phi \choose j+k}{N_\Phi \choose j-k}} \tilde
p_{m,j}(k)\,,
\end{equation}
\end{subequations}
where the $\tilde p_{m,j}(k)$ are polynomials different from the
$p_{m,j}(k)$, but with the same general properties noted for the latter.
The equivalence of Eqs \eqref{etasphere0} and \eqref{etasphere} will be
explained in Appendix \ref{appA}, where a recursive definition of the
polynomials  $\tilde p_{m,j}(k)$ is also given.

For the cylinder, we will work in a Landau gauge where the vector
potential is independent of $y$, and we impose periodic boundary
conditions in $y$ with period $L_y$.\cite{haldanerezayi94} Two-particle
wave functions are then of the form $\psi(z_1,z_2)=f(z_1,z_2) e^{-\frac
12(x_1^2+x_2^2)}$, where $f(z_1,z_2)$ is holomorphic and periodic in
$y$, i.e., $f(z_1,z_2)=f(z_1+iL_y,z_2)=f(z_1,z_2+iL_y)$. The
pseudopotential $V_m$ as defined for the disk does not respect this
boundary condition. One must therefore work with ``periodized'' versions
of these pseudopotentials. For this we may view the full pseudopotential
as a sum over particle pairs of  the Landau level projected version of
an ultra-short ranged pair potentials $V_m(z_1-z_2)$, e.g.,
$V_1(z_1-z_2)=\hat{P}_{\sf LLL}\nabla^2 \delta(z_1-z_2)\hat{P}_{\sf
LLL}$,\cite{Trugman} and regard the cylinder-version of this potential
as $V^{\sf cyl}_m(z_1-z_2)= \sum_\ell V_m(z_1-z_2+i\ell L_y)$. Here
$\hat{P}_{\sf LLL}$ is the projection onto the LLL. Moreover, we note that
$V_m\psi(z_1,z_2)$,  where $\psi$ satisfies the periodic boundary
conditions defined above, is still periodic under {\em simultaneous}
shifts of $z_1$ and $z_2$ by $iL_y$, since $V_m$ acts only on the
relative coordinate. We may thus write $V_m^{\sf cyl}\psi(z_1,z_2)=
{\cal P} V_m\psi(z_1,z_2)$, where ${\cal P} f(z_1,z_2)= \sum_\ell
f(z_1+i\ell L_y, z_2)$. 

From these considerations, it follows that
$V_m^{\sf cyl}$ projects onto the subspace of wave functions of the
following form,
\begin{eqnarray}\label{Vcyl}
&&  V_m^{\sf cyl} \psi(z_1,z_2)= \\ \nonumber
&& \sum_{\sf{\ell}} a(R+\frac{{\sf \ell}}{2} iL_y) (z+i{\sf \ell} L_y)^m
e^{\frac{1}{8}(z+i{\sf \ell} L_y)^2} e^{-\frac 12 x_1^2-\frac 12 x_2^2},
\end{eqnarray}
where $R=(z_1+z_2)/2$, $z=z_1-z_2$,  $a(R)$ is a holomorphic function
satisfying the periodicity $a(R+iL_y)=a(R)$, and the ${\sf \ell}=0$ term
is just  $V_m \psi(z_1,z_2)$. The first exponential is picked up by
first going to the symmetric gauge, there evaluating the effect of
$V_m$, and then transforming back to Landau gauge. It is worth noting
that unlike $V_m \psi(z_1,z_2)$, $V_m^{\sf cyl} \psi(z_1,z_2)$ does not
in general have an $m$th order zero as $z_1\rightarrow z_2$. On the
other hand, what matters is that any $\psi(z_1, z_2)$ satisfying $V_m
\psi(z_1,z_2)\equiv 0$ also satisfies $V_m^{\sf cyl} \psi(z_1,z_2)\equiv
0$, since the first condition is equivalent to $a(R)\equiv 0$. Moreover,
the converse is also true, as one may verify by a Wick rotation in both
$z_1$, $z_2$ of the holomorphic part of \Eq{Vcyl} and subsequent Poisson
resummation. From this it is not difficult to show via induction in $m$
that states satisfying $V_k^{\sf cyl} \psi(z_1,z_2)\equiv 0$ for all
$0\leq k\leq m$, with $k$ being odd (for bosons) or even (fermions), must
have at least an $(m+2)$th order zero as $z_1\rightarrow z_2$, just as
it is in the other geometries.
Equation \eqref{Vmsec} still applies with $H_{V_m^{\sf cyl}}$ in place
of $H_{V_m}$. The indices on ladder operators now refer to the momentum
about the cylinder axis of the orbitals they create/annihilate, in units
of $\kappa\equiv 2\pi/L_y$. It follows from this that $H_{V_{m,j}^{\sf
cyl}}$ now projects onto states of the form \eqref{Vcyl} with
$a(R)=\exp(\kappa j (z_1+z_2))$. 
Since this defines a one-dimensional subspace, the arguments given above
for the disk and sphere still apply, and $M^j_{kl;m}=\eta_{k}(j,m)
\eta_{l}(j,m)$. Note that straightforward ``periodization''  of the 
pseudopotential preserves
the normalization  \eqref{etanorm} only in the thermodynamic limit.
However, on the cylinder the $\eta_{k}$ are truly independent of $j$ due
to translational invariance.

For completeness, we finally give a formula for the $\eta_{k}$ in
the cylinder geometry. 
We begin by writing Haldane's formula for the
operator $P_m({\sf ij})$ in disk geometry:\cite{Haldane}
\begin{eqnarray}\label{Pmij}
    P_m({\sf ij})= \frac{1}{\pi} \int \int dq_x dq_y\,
     L_m[q^2]e^{-\frac{q^2}{2} + i{\mathbf{q}}\cdot({ \mathbf{\tilde
     r}}_{\sf i}-{ \mathbf{\tilde r}}_{\sf j})}\,.
\end{eqnarray}
Here, $L_m$ is the $m$th Laguerre polynomial, and $\mathbf{\tilde
r}_{\sf i}=({\tilde x}_{\sf i},{\tilde y}_{\sf i}) =\hat{P}_{\sf
LLL}(x_{\sf i},  y_{\sf i})\hat{P}_{\sf LLL}$ is the projected position
or ``guiding center'' of the $\sf i$th particle. According to the above
discussion, we can use this expression  for the cylinder after the
replacement $\int dq_y\rightarrow \sum_{q_y}$ where $q_y$ is quantized
in multiples of $\kappa=2\pi/L_y$. $\tilde x$ is quantized in the same
manner, which follows from the commutation relation $[\tilde x, \tilde
y]=i$, together with the fact that $\tilde y$ acquires angular
character, $\tilde y\equiv \tilde y+L_y$. The operator $c_r^\dagger$
creates an eigenstate of $\tilde x$ with eigenvalue $\kappa r$.
Restricting ourselves to two particles for the moment, the operator
$\exp(iq_x(\tilde x_1-\tilde x_2))$ is diagonal in the basis
$c^\dagger_{r_1}c^\dagger_{r_2}|0\rangle$, whereas the operator
$\exp(iq_y(\tilde y_1-\tilde y_2))$ shifts the $\tilde x$ eigenvalue of
particle $1$ (particle $2$) by $-q_y$ (by $q_y$). These observations
lead to the identification of the operator $\sum_\pm e^{\pm
i{\mathbf{q}}\cdot({ \mathbf{\tilde r}}_{\sf i}-{ \mathbf{\tilde
r}}_{\sf j})}$ with the second quantized operator $\sum_\pm
\sum_{r_1,r_2}  e^{-iq_x q_y}  e^{iq_x \kappa (r_1-r_2)}  c_{r_2 \mp
q_y}^\dagger c_{r_1\pm q_y}^\dagger c^{\;}_{r_1}c^{\;}_{r_2}$,  where
the first exponential comes from an application of the
Baker-Hausdorff-Campbell identity in rearranging the exponential of
non-commuting operators as a product of two exponentials. In Eq.
\eqref{Pmij}, and comparing with Eq. \eqref{Vmsec}, this yields the
identification of $M^j_{xy;m}=\eta_{x}(j,m) \eta_{y}(j,m)=\eta_x \eta_y$
with
\begin{eqnarray}\label{etaxy}
  \eta_{x}\eta_{y} =\;&&\frac{2}{\pi} \int_{-\infty}^\infty dq\,
L_m[q^2+(x-y)^2]\nonumber\\
  &&\hspace*{0.5cm} \times\, e^{{-\frac 12 (q^2+(x-y)^2) +iq(x+y)}}\,,
\end{eqnarray}
where we have restricted ourselves to the case $\kappa=1$. The general
case is obtained by letting $x\rightarrow \kappa x$ and similarly for
$y$. One can write a compact expression for Eq. \eqref{etaxy} in terms
of Hermite polynomials  of even order
\begin{eqnarray}\label{etaxyHerm}
  \eta_{x}\eta_{y} =\;&&\frac{e^{-(x^2+y^2)}}{2^{2m-3/2}\sqrt{\pi}\,
  m!} \sum_{l=0}^m (-1)^l \binom{m}{l} \,  \nonumber\\
  &&\hspace*{0.5cm} \times\, H_{2m-2l}[x+y] \, H_{2l}[x-y] \, ,
\end{eqnarray}
by using the series expansion
\begin{eqnarray}
 L_m[q^2+(x-y)^2]=\;&&\frac{(-1)^m}{2^{2m}\, m!} \sum_{l=0}^m \binom{m}{l} 
\,  \nonumber\\
  &&\hspace*{0.2cm} \times\, H_{2m-2l}[q] \, H_{2l}[x-y] \, ,
\end{eqnarray}
and performing integration over $q$. 
The above expression (\ref{etaxyHerm}) readily simplifies once we 
transform to an $X,Y=(x \pm y)$ coordinate frame,  express the Hermite 
polynomials in Eq. (\ref{etaxyHerm}) via standard creation operators 
$a_{X,Y}^{\dagger}$ acting  on Gaussians in $X,Y$, whence this 
reduces to $[(a_{X}^{\dagger})^{2} - (a_{Y}^{\dagger})^{2}]^{m}$ or,
equivalently,   $[4 \tilde{a}^{\dagger}_{x}
\tilde{a}^{\dagger}_{y}]^{m}$, with $\tilde{a}^{\dagger}_{x}$ scaling as
$(\frac{1}{2} \partial_{x} -x)$, acting on  a Gaussian in $x$ multiplied
by a Gaussian in $y$. In the aftermath, the right hand side of Eq.
(\ref{etaxyHerm}) factorizes into decoupled functions in $x$ and $y$ as
the left hand side implies with
\begin{eqnarray}\label{etaxyHermn}
  \eta_{x} =\frac{e^{-x^2}}{2^{\frac{m}{2}-\frac{3}{4}} \,
\sqrt{m!\sqrt{\pi}}} \ H_m[\sqrt{2} \, x] \, ,
\end{eqnarray}
which are given in Table \ref{etatable} for $m=0,\cdots,7$. 
Note that Eq. \eqref{etaxyHermn} simplifies the expression 
recently given in Ref. \onlinecite{thomale}, where higher-body
pseudopotentials on the cylinder were also treated.

For the torus geometry, similar arguments could be made for the
separated form of the pseudopotentials. Instead, we refer to the
relation between the second quantized forms of these  potentials for the
cylinder and torus geometries given in Section \ref{equivalence}. 
 
\begin{table}[tbh]
\begin{tabular}{p{2.2cm}  p{6.2cm}}
\hline\hline
$m$ & $\eta_{x} \times (\pi/8)^{1/4}\exp[{x^2}]$  \\ [0.5ex] 
\hline\\
$0$ & $1$  \\[0.5ex] 
$1$ & $2x$   \\[1ex] 
$2$ & $\sqrt{1/2}\,(4x^2-1)$ \\[1ex] 
$3$  & $\sqrt{2/3}\,x(4x^2-3)$  \\[1ex]
$4$ & $\sqrt{1/24}\,(16x^4-24x^2+3)$  \\[1ex] 
$5$ & $\sqrt{1/30}\, x(16x^4-40x^2+15)$   \\[1ex] 
$6$ & $\sqrt{1/720}\,(64x^6-240x^4+180x^2-15)$ \\[1ex] 
$7$ & $\sqrt{1/1260}\,x(64x^6-336x^4+420x^2-105)$ \\[1ex] 
\hline
\end{tabular}
\caption{\label{etatable}
The polynomial parts of the coefficients $\eta_{x}$  defining the
$m$th Haldane pseudotential in second quantization as in Eq.
\eqref{V1QHE} for a cylinder with $\kappa=1$. For general cylinders,
$\eta_x$ can be obtained from the given expressions via substitution
$x\rightarrow \kappa x$ and overall multiplication by $\sqrt{\kappa}$.}
\end{table}

\subsection{Quantum Hall algebra}
\label{qha}

We are interested in identifying the algebra of interactions relevant
for the QH problem. Define the operators
\begin{eqnarray}\label{Tj}
T_{j s;m}^+ &=& \!\! \sum_{k(j)} \eta_{k}^{2s-1} c^\dagger_{j+k}
c^\dagger_{j-k} \ , \ T_{j s;m}^- =  \!\!  \sum_{k(j)} \eta_{k}^{2s-1}
c^{\;}_{j-k} c^{\;}_{j+k} \nonumber \\
T_{j s;m}^z&=& \frac{1}{2}  \sum_{k(j)}  \eta_{k}^{2s} (n_{j+k}+n_{j-k} -1) \ ,
\end{eqnarray}
where $s \in \mathbb{Z}$, and the number operator is defined as
$n_{j+k}=c^{\dagger}_{j+k} c^{\;}_{j+k}$ and, as throughout, the sum is
performed over $k$ following the convention of Eq. (\ref{ksum0}).  These
operators close an infinite-dimensional affine Lie algebra without a
central extension
\begin{eqnarray}\label{affine}
{[}T_{j s;m}^+ ,  T_{j s';m}^- ]&=& 2 \, T_{j (s+s'-1);m}^z \nonumber \\
{[}T_{j s;m}^z ,  T_{j s';m}^\pm ]&=& \pm \, T_{j (s+s');m}^\pm  .
\end{eqnarray}
With these, the lowest order Haldane pseudopotential of Eq.
(\ref{V1QHEtor}) becomes
\begin{eqnarray}
H_{V_{1}} &=& \sum_{0 < j < L-1} T^+_{j 1;1}T^-_{j 1;1},
\end{eqnarray} 
which explicitly displays its positive semi-definite character
($g_{1}>0$ in Eq. (\ref{sumQH})). For this special
case, it is then known that the Hamiltonian has zero energy ground
states at filling fraction $\nu\leq 1/3$, as we pointed out above and
will be analyzed in more detail in later sections. 

Far more generally, a {\it generic LLL  QH Hamiltonian} of the form of
Eq. (\ref{sumQH}) can be written as
\begin{eqnarray}
\hspace*{-0.7cm}\label{HLLL}
\widehat{H}_{\sf QH}&=& \sum_{m} g_{m} \sum_{0 < j < L-1} T^+_{j
1;m}T^-_{j 1;m} .
\end{eqnarray}
Now here is an important point whose meaning will become clear in future
sections: within each sector of fixed $m$ and $j$, the argument in the
sum of Eq. (\ref{HLLL}) is of the form of an exactly-solvable  
RG pairing Hamiltonian
(Eq. (\ref{HamiltG})). 

\subsection{Topological aspects of the Quantum Hall  problem: An exact
equivalence of the disk,  cylinder, and spherical  geometries}
\label{equivalence}

In the following,  we will be interested in the task of characterizing zero
energy states or ``zero modes''. To make the discussion lucid we will 
concentrate on the zero modes of the $H_{V_1}$
pseudopotential Hamiltonian. As we will discuss in Section \ref{nullspace},
the latter are constrained by the condition
\begin{eqnarray}\label{zeromode}
 T^-_{j1;1} |\Psi\rangle=0
\end{eqnarray}
for all $j$, where $T^-_{j1;1}$ is defined in terms of the parameters
$\eta_k$ given in Table \ref{table:nonlin} for various geometries.
Despite the different appearance of these coefficients, 
the tasks of finding the zero energy eigenstates (zero modes) of $V_1$
for the disk, cylinder, and sphere are {\it exactly} equivalent, for
which we will now provide appropriate transformations in second
quantization. It is intuitive that such transformations exist, as it is well
appreciated that universal features of topologically ordered
states are insensitive to geometric details, and only depend
on the genus number (number of handles) of the system.\cite{wen} Such universal features
do not generically include the counting of zero modes (at filling factors below the incompressible one). However,
for ``fixed point''  Hamiltonians such as the parent Hamiltonians of the
Laughlin states, this is the case. At the first quantized
level, this is a manifestation of the polynomial structure 
wave functions display for the disk/cylinder/sphere geometries, which has been used
extensively in the derivation of counting formulas for zero modes, both for the
$V_1$ pseudopotential as well as other parent Hamiltonians. \cite{RR, ardonne} 

\begin{figure}[htb]
\includegraphics[width=0.99\columnwidth]{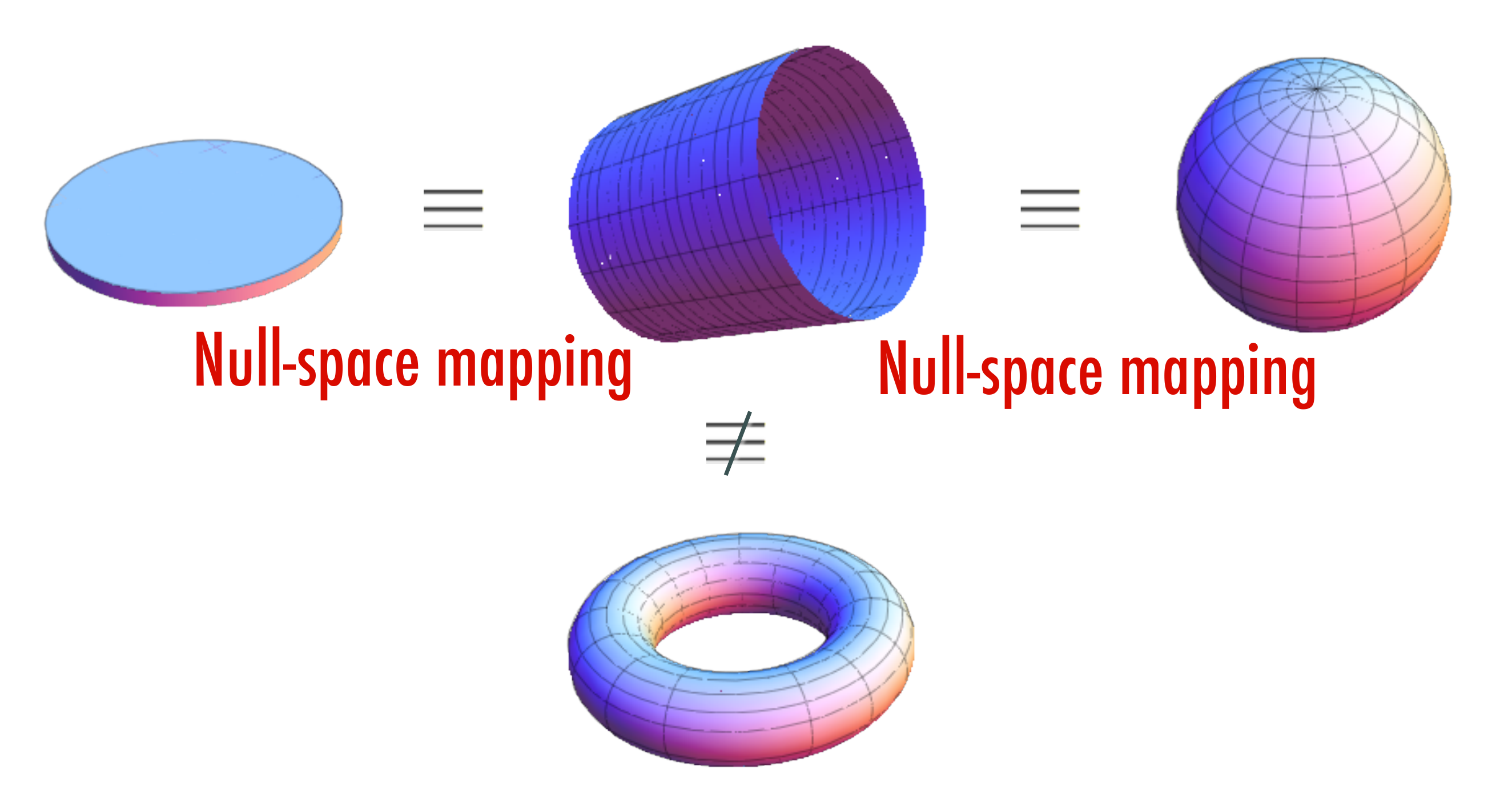}
 \caption{Topological equivalence between geometries sharing the same 
 genus number. }
 \label{TopoEq.fig}
\end{figure}

Note that the disk, cylinder, and sphere all have vanishing genus
number  (while the genus number of the torus is $1$).  
Below we will show how the equivalence between zero modes
for these different geometries is recovered in second quantization.
As
evident from Table \ref{table:nonlin}, the generic structure of LLL
orbitals in these geometries is
\begin{eqnarray}
    \phi_r= {\cal N}_r \ \xi^r \times (r{\sf -independent~ function}), 
\end{eqnarray}
where $\xi=z$, $e^{\kappa z}$, $u/v$ for the disk, cylinder, and sphere,
respectively, is a holomorphic factor, with 
$u=e^{-i\frac{\varphi}{2}}\sin(\frac{\theta}{2})$,
$v=e^{i\frac{\varphi}{2}}\cos(\frac{\theta}{2})$, and  ${\cal N}_r$ is a
geometry-dependent normalization factor. General LLL wave functions are
thus polynomials in $\xi$. Note that for a cylinder with inverse radius
$\kappa\rightarrow 0$, we have ${\cal N}_r\equiv 1$. Consider now the
similarity transformation that acts via
\begin{eqnarray}
   c^{\;}_r \rightarrow S c^{\;}_r S^{-1}={\cal N}_r c^{\;}_r,
\end{eqnarray}
where ${\cal N}_r$ corresponds to any given geometry. We can think of
this transformation as changing the normalization conventions of
polynomials for the $\kappa=0$ cylinder to that of any other geometry.
Specifically, we may take $S=S_{\sf c}\equiv\exp[\frac 12 \kappa^2
\sum_r r^2 c_r^\dagger c^{\;}_r]$  for a cylinder at finite $\kappa$,
$S=S_{\sf d}\equiv \exp[\frac 12 \sum_r \ln(2^r r!)c^\dagger_r
c^{\;}_r]$ for the disk, and $S=S_{\sf s}\equiv\exp[-\frac 12 \sum_r
\ln{N_\Phi \choose r}c^\dagger_r c^{\;}_r ]$ for the sphere. Let us
denote by $t^+_{j}$, $t^-_{j}$, $t^z_{j}$ the operators defined in Eq.
(\ref{Tj}) with $\eta_k\equiv k^{m=1}$ ($t^\zeta_{j}=t^\zeta_{j;m=1}$, with 
$\zeta=\pm,z$).  Equation \eqref{zeromode} with $T^-_{j1;1}=t^-_{j}$
then corresponds to the $\kappa=0$ cylinder.
It is further easy to verify that the  operators $T^-_{j1;1}$ for a general
cylinder, a disk, or a sphere are then obtained via 
\begin{eqnarray}\label{1to1}
    T^-_{j1;1}= f_j S t^-_{j} S^{-1}\,,
\end{eqnarray}
see Fig. \ref{TopoEq.fig}, where $f_j$ is a positive factor that depends
on the geometry and $j$, and $S$ depends on the geometry as shown
above.  Therefore, if $|\Psi\rangle$ satisfies $t^-_{j}|\Psi\rangle=0$,
then $S|\Psi\rangle$ satisfies Eq. (\ref{zeromode}), i.e., $T^-_{j1;1}
S|\Psi\rangle=0$. It is thus sufficient in principle to work in the
$\kappa=0$ cylinder geometry, and study the zero modes of the operators
$t^-_{j}$. Note that we could always obtain the coefficients $f_j$ from
the condition \eqref{etanorm} if desired.

We caution that for higher $m$, the one-to-one correspondence between 
pseudopotentials in different geometries  ceases to hold in the strict
sense of \Eq{1to1}. The reason for this is that states related by the
transformations defined above generally correspond to the same
polynomials in the first quantized description for the respective
geometries.
For $m>1$, however, the rank 1 projectors $H_{V_{m,j}}$ of Eq.
\eqref{Vmsec} will in general project onto states having a different
polynomial structure for the different geometries.
On the other hand, we still have the following statement: 
For fixed $j$, the transformed states $St^+_{j;m}|0\rangle$ for $0\leq
m\leq {\sf M}$ (and $m$ even/odd for bosons/fermions here and below)
span the same subspace as the states ${T}^+_{j1;m}|0\rangle$ defined in
\Eq{jmstate}.
Here, the ${T}^+_{j1;m}$ and the transformation $S$ refer to the same
geometry.
The reason for this is that in any geometry, the $\eta_k$'s are
proportional to an $m$th order polynomial in $k$ with parity $(-1)^m$
(see Eqs.  \eqref{etadisk}, \eqref{etasphere}, and \eqref{etaxyHerm}),
and the $\eta_k$ defining $t^+_{j;m}$ are just equal to $k^m$. All other
$k$-dependent factors are independent of $m$ and are taken care of by
the transformation $S$. Within the two-particle sub-space, the common null space of the operators
$H_{V_{m,j}}={T}^+_{j1;m} {T}^-_{j1;m}$, $0\leq m\leq {\sf M}$, is just
the space orthogonal to all the states ${T}^+_{j1;m}|0\rangle$, and
similarly the common null space of the ``transformed''  operators  $(S
t^-_{j;m} S^{-1})^\dagger S t^-_{j;m} S^{-1}$, $0\leq m\leq {\sf M}$ is
the space orthogonal to all the states $St^+_{j;m}|0\rangle$, $0\leq
m\leq {\sf M}$. Thus in any geometry,  the operators
$H_{V_{m,j}}$, for $0\leq m \leq {\sf M}$, always have the same {\em
common} null space as the operators $(S t^-_{j;m} S^{-1})^\dagger S
t^-_{j;m} S^{-1}$. This statement immediately carries over from the two-particle
subspace to the full Fock space, since the operators in question are two-body operators.
This will be of some importance in Section \ref{2ndhole}.

Note that the above equivalence of null spaces holds for fixed particle number
{\em and} number of Landau level orbitals. Working with a finite number
of orbitals requires ``hard orbital cutoffs'' in the cylinder and disk geometries,
but is the usual situation for the sphere. There is thus no contradiction
with the common knowledge that these three geometries have different
numbers of edge modes. Edge modes are present, in particular,
for the usual infinite or half-infinite
orbital lattice associated to the cylinder or disk geometry, respectively.
Conversely, however, edge modes {\em can}  be present in the spherical geometry as well,
if, say, we populate only the northern half with a FQH state having an edge at the equator.

We finally observe that if $\eta_{k}^{\sf cyl} $  corresponds to the
pseudopotential $V_m$  on the cylinder, then for the torus it can
always be obtained by further periodizing the cylinder. This can be done
directly in second quantization:
\begin{eqnarray}
 \eta_{k}^{\sf tor}=\sum_\ell \eta^{\sf cyl}_{k+\ell L}\,.
\end{eqnarray}
Here and in Table \ref{table:nonlin}, we restrict ourselves to tori with
purely imaginary modular parameter $\tau=iL_y/L_x=2\pi i/\kappa^2 L$,
where we introduced $L_x=\kappa L$.


\section{Strongly-coupled states of matter}
\label{strong}

In this section, we relate the QH Hamiltonian of Section \ref{2nd_quant}
to the hyperbolic RG type models encountered,  for instance,  in the
study of $(p_{x}+ip_{y})$ superconductors. In particular, in Section
\ref{exact} we  demonstrate that within each sector of fixed $m$ and
$j$,  the Hamiltonian of Eq. (\ref{HLLL}) represents a new
exactly-solvable  model, that we call QH-RG,  which belongs precisely to
the hyperbolic RG class. We then examine (Section \ref{Hilbert}) the
Hilbert space dimension associated with the QH-RG problem. In each such
sector of fixed $j$ and $m$, the spectral problem can be determined via
Bethe Ansatz  as we explicitly demonstrate in Section
\ref{eigenspectrum}.  We conclude our analysis of the QH-RG Hamiltonian 
in this section, by highlighting the symmetry properties of the RG
equations (Section \ref{sym_prop}). The full problem  formed by the sum
of all (generally non-commuting) QH-RG Hamiltonians will be investigated
in Section \ref{full_QH}. 

\subsection{An exactly-solvable model}
\label{exact}

The XXZ Gaudin algebra \cite{NuclPhysB} is an affine Lie algebra
generated by  operators ${\sf S}^z(x), {\sf S}^\pm(x)$ with  commutation
relations
\begin{eqnarray}
  \left[ {\sf S}^z(x), {\sf S}^\pm(y) \right]  &=& \pm \left( X(x,y)
 {\sf S}^\pm(x) -Z(x,y) {\sf S}^\pm(y) \right), \nonumber \\
  \left[ {\sf S}^+(x), {\sf S}^-(y) \right]  &=& 2 X(x,y) \left( {\sf
S}^z(x) - {\sf S}^z(y) \right),
 \label{eq:gaudinalgebra}
\end{eqnarray}
in terms of anti-symmetric functions $X(x,y)$ and $Z(x,y)$ satisfying
the following condition for all $x$, $y$ and $z$
\begin{equation}
  \left[ Z(x,y)-Z(x,z) \right] X(y,z) - X(x,y) X(x,z) = 0.
  \label{eq:gaudineq}
\end{equation}

A representation of the XXZ Gaudin algebra in terms of a number ${\cal C}(j)$ 
(see Eq. \eqref{numbkl}) of
$su(2)$ spins, $\{S^z_{jk}, S^{\pm}_{jk}\}$, labeled by the (in
principle arbitrary) quantum numbers $j \in [j_{\sf min}, j_{\sf max}]$ and 
$k\in[k_{\sf min}, k_{\sf max}]$ is given by
\begin{eqnarray}
\label{su2s}
  {\sf S}_j^z(x) &=& -\frac{1}{2} - \sum_{k(j)} Z(x,\eta_{k}) S^z_{jk},
 \nonumber \\ 
 {\sf S}_j^\pm(x) &=& \sum_{k(j)} X(x,\eta_{k}) S^{\pm}_{jk},
\end{eqnarray}
with $\eta_{k}$ being {\it arbitrary parameters} (eventually we will equate these general parameters 
to be the very same constants $\eta_{k}(j,m)$ that appeared in our decomposition of the Haldane 
pseudopotentials). In this  representation one
can define a set of ${\cal C}(j)$ linearly independent constants of
motion, which commute among themselves
\begin{eqnarray}
  R_{jk} &=& S^z_{jk} - \sum_{l(j),l\neq k} 
  X(\eta_{k},\eta_{l}) \left( S^+_{jk} S^-_{jl} + S^-_{jk} S^+_{jl} \right)
	\nonumber \\
       &-& 2 \sum_{l(j),l\neq k} Z(\eta_{k},\eta_{l}) S^z_{jk}
S^z_{jl},
  \label{eq:integrals_of_motion}
\end{eqnarray}
Linear combinations of these operators allow for the construction  of an exactly solvable
RG Hamiltonian
\begin{eqnarray}
  H_{{\sf G}j} &&= \sum_{k(j)} \epsilon_{k} S^z_{jk} \nonumber \\ 
  && - \sum_{k(j),l(j)} (\epsilon_{k}-\epsilon_{l}) X(\eta_{k},
	\eta_{l}) S^+_{jk} S^-_{jl} \nonumber \\
  &&- \sum_{k(j),l(j)} (\epsilon_k-\epsilon_l) Z(\eta_k,\eta_l) S^z_{jk}
S^z_{jl}.
  \label{eq:rg_hamiltonian}
\end{eqnarray}
The eventual $m$ and $j$ dependence of $H_{{\sf G}j}$ stems from that of
the generators in Eq. (\ref{su2s}). This generic RG Hamiltonian commutes
with the squared spin operators ${\bf
S}_{jk}^2=S^z_{jk}(S^z_{jk}-1) + S^+_{jk}S^-_{jk}$, and with the total
spin operator ${\bf S}_j= \sum_{k(j)} {\bf S}_{jk}$.

A consequence of the Jacobi identity, Eq. (\ref{eq:gaudineq}), is that
\begin{equation}
  X(x,y)^2 - Z(x,y)^2 = \Gamma,
  \label{eq:gaudingamma}
\end{equation}
where $\Gamma$ is a constant independent of $x$ and $y$. In this work we
will focus on the properties of the {\em hyperbolic} RG model, which
correspond to $\Gamma =-\bar{g}^2/4 <0$.  It is interesting to mention
that the $p_x+i p_y$ integrable pairing model  belongs to this class. 
\cite{Romb}   Any set of functions $X(x,y)$ and $Z(x,y)$ that fulfills
Eqs. (\ref{eq:gaudineq}) and (\ref{eq:gaudingamma}),  can be mapped onto the
following parameterization \cite{NuclPhysB}
\begin{equation}
 X(x,y) = -\bar{g} \frac{x y}{x^2-y^2}, \ \ \
 Z(x,y) =- \frac{\bar{g}}{2} \frac{x^2+ y^2}{x^2 - y^2}.
\end{equation}

Using this parameterization, setting $\epsilon_k = \lambda_j \eta^2_{k}$,
and subtracting a diagonal term $ -\bar{g} \sum_{k(j)}\eta^2_k {\bf
S}^2_{jk}$, one obtains an interesting form for the Hamiltonian of Eq.
(\ref{eq:rg_hamiltonian})
\begin{eqnarray}
H_{{\sf G}j}&=&\lambda_j(1+\bar{g}(S_j^z-1)) \sum_{k(j)} \eta_k^2
S^z_{jk} \nonumber \\
&&+\lambda_j \bar{g} \sum_{k(j),l(j)} \eta_k \eta_l S^+_{jk} S^-_{jl} ,
\label{modlppip}
\end{eqnarray}
where for a {\it fixed} $j$, $j\pm k , j\pm l$ are all non-negative
integers in the interval $[0,L-1]$ as before. The  parameters
$\lambda_j, \bar{g}$ (which can be positive or negative) and $\eta_k$
are arbitrary in principle.

A possible fermionic representation of the $su(2)$ spin {\it algebra},
similar to  the $p_x+ip_y$ superconductor
\cite{Romb} is given by
\begin{eqnarray}
S_{jk}^+ &=& c^\dagger_{j+k} c^\dagger_{j-k} \ , \
S_{jk}^- =c^{\;}_{j-k} c^{\;}_{j+k} \nonumber \\
S_{jk}^z&=& \frac{1}{2} (n_{j+k} + n_{j-k}-1) .
\end{eqnarray}

\begin{figure}[htb]
\includegraphics[width=0.99\columnwidth]{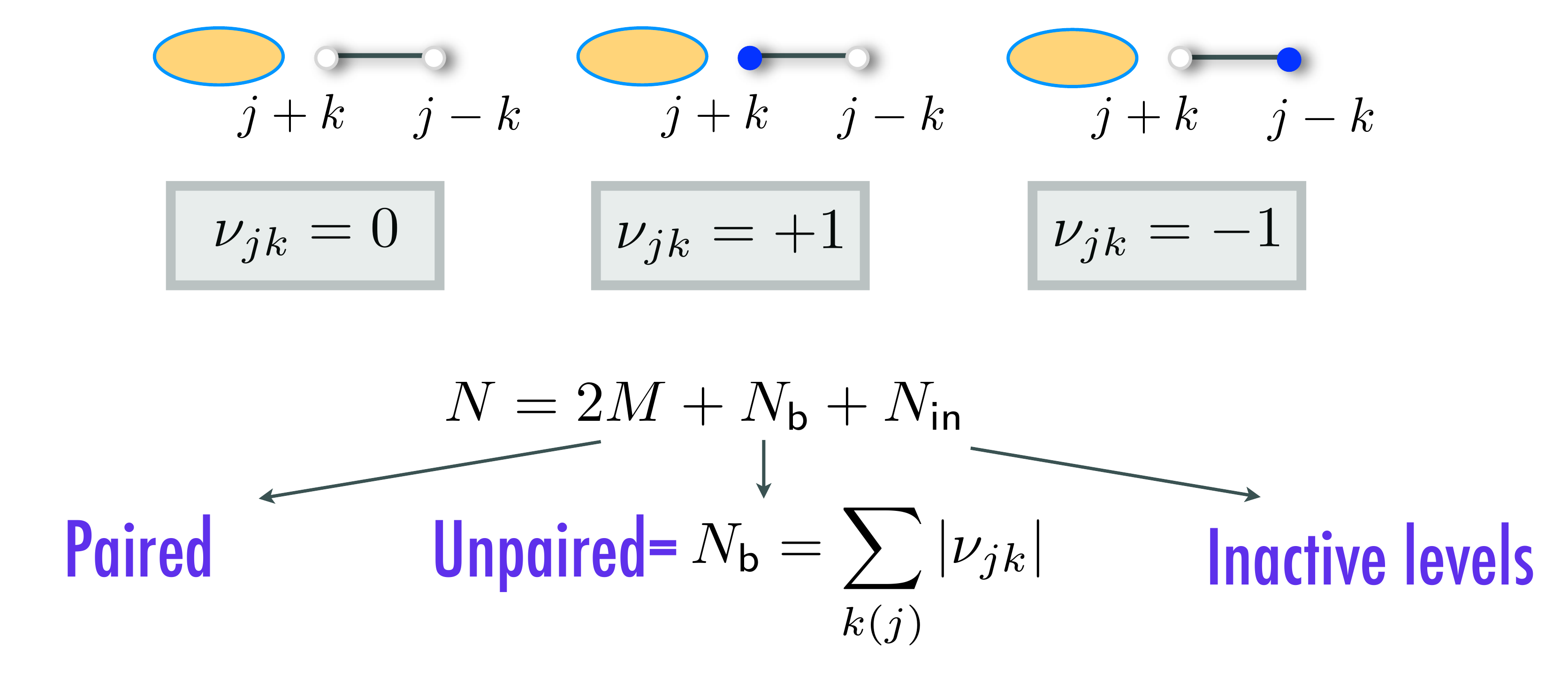}
 \caption{Possible electronic configurations for active level ${\bf k}\equiv[j+k,j-k]$, 
  and corresponding values of their seniority $\nu_{jk}$. Dark circles correspond to 
  an occupied orbital in $\ket{\{\nu_{jk}\}}$, while white ones to unoccupied ones. Ellipsis correspond 
  to pairs in active levels (i.e., in ${\sf S}_j^+$). Notice that configurations with $\nu_{jk}=\pm 1$ 
  Pauli block the corresponding level ${\bf k}$ in ${\sf S}_j^+$. Thus, 
  active levels form either  pairs or are unpaired.  }
 \label{Senio.fig}
\end{figure}

As mentioned above, the value of $j$ is arbitrary in the interval 
$[\frac{1}{2},L-\frac{3}{2}]$ (see Eq. \eqref{jvalues}). However, once the value of $j$ is chosen,
it  classifies completely the basis states into an active space of
${\cal C}(j)$  active levels  ${\bf k} \equiv [j+k,j-k]$,  and a set of
inactive levels  $\{i_1,i_2,\cdots, i_{L-2 {\cal C}(j)} \}$ which
includes the remaining  $L-2 {\cal C}(j)$ levels left out of the  active
set. This classification allows us to define an $su(2)$ vacuum {\em
state} $\ket{\nu(j)}$, which is annihilated by  the lowering operators
$S_{jk}^-\ket{\nu(j)}=0 $ as
\begin{eqnarray}
\ket{\nu(j)}\equiv\ket{\{ \nu_{jk}\}} \otimes \ket{\nu_{\sf in} }.
\end{eqnarray}
The {\it seniorities} $\nu_{jk}$ are defined as follows: $\nu_{jk}=0$ if the level  
${\bf k}$ is empty or doubly occupied (not in the vacuum $\ket{\{ \nu_{jk}\}} $), 
and $\nu_{jk}= \pm 1$ if there is a
single electron with momentum $j\pm k$  (see Fig. \ref{Senio.fig}). 
The two different non-zero
values  for the seniorities $\nu_{jk}$ are associated with a spin 1/2
degree of freedom. 

The state $\ket{\nu_{\sf in}}$ defines a configuration of $N_{\sf in}$
electrons distributed among the $L-2 {\cal C}(j)$ inactive levels 
\begin{eqnarray}
\ket{\nu_{\sf in}}\equiv c^\dagger_{i_1}c^\dagger_{i_2} \cdots \,
c^\dagger_{i_{N_{\sf in}}} \ \ket{0} .
\end{eqnarray}
Therefore, the  vacuum is an eigenstate of the associated operator
$S_{jk}^z$ 
\begin{eqnarray}
S_{jk}^z\ket{\nu(j)}=\frac{1}{2} (|\nu_{jk}|-1)\ket{\nu(j)} \equiv
-s_{jk}\ket{\nu(j)}.
\label{Szjk}
\end{eqnarray}

Additional symmetries become manifest in the fermionic language. In
particular,  the $su(2)$ algebra
\begin{eqnarray}
\tau_{jk}^+ &=& c^\dagger_{j+k} c^{\;}_{j-k} \ , \
\tau_{jk}^- =c^{\dagger}_{j-k} c^{\;}_{j+k} \nonumber \\
\tau_{jk}^z&=& \frac{1}{2} (n_{j+k} - n_{j-k}) ,
\label{gaugesym}
\end{eqnarray}
generates the gauge symmetry of $H_{{\sf G}j}$ responsible for Pauli
blocking, \cite{NuclPhysB} with seniority $\nu(j)$ representing a good
quantum  number. This gauge symmetry is no longer a  symmetry of 
pseudopotential Hamiltonians  $H_{V_m}$ formed by the sum of the
individual Hamiltonians $H_{{\sf G}j}$ for each value of $j$.  
Similarly, the total angular momentum operator defined as  ($r$ is an
integer)
\begin{equation}
\hat{J}=\sum_{r=0}^{L-1} r \ n_r ,
\end{equation}
is also a symmetry of the Hamiltonian $H_{{\sf G}j}$. It is easy to
check that $[\hat{J},S_{jk}^\pm]=\pm2j S_{jk}^\pm$, implying that the
angular momentum of each pair is $2j$ and that the maximum possible angular
momentum of each electron in the pair is also $2j$. Moreover, the state
$j$ does not participate in pairing, and  it can be empty or  occupied
by a single electron, defining a seniority $\nu_{j0}=0,1$, respectively.

Assume that the total number of electrons, a good quantum number,  is $N
=2 M+N_{\nu(j)}$, where $M$ is the number of pairs with angular momentum
$2j$, and $N_{\nu(j)}=N_{\sf b}+N_{\sf in} $ is the total number  of
unpaired electrons with
\begin{eqnarray}
N_{\sf b}=\sum_{k(j)} |\nu_{jk}|=\sum_{k(j)} (1-2s_{jk}) .
\end{eqnarray}
Note that the total number of electrons has three contributions: 
(i) the $2M$
electrons that participate in the pair mechanism, and the unpaired
electrons  $N_{\nu(j)}$ which in turn are split into (ii) $N_{\sf b}$
electrons blocking active  levels (see the two $\nu_{jk} = \pm 1$ cases shown in Fig. \ref{Senio.fig}) 
and (iii) $N_{\sf in}$ electrons distributed among inactive levels. 
Then a seniority
configuration of  $N_{\nu(j)}$ unpaired electrons
\begin{eqnarray}
[j+k_{1}, j+k_{2},\cdots, j+k_{N_{\sf b}};{{i_1},{i_2},\cdots,{i_{N_{\sf
in}}}}] ,
\end{eqnarray}
is an eigenstate of $H_{{\sf G}j}$ with $M$ pairs, $\ket{\Phi_{M\nu(j)}}$,
satisfies
\begin{eqnarray}
\hspace*{-0.7cm}
S^z_j\ket{\Phi_{M\nu(j)}}&=&\frac{1}{2} \left(2M+N_{\nu(j)}-{\cal C}(j)
\right) \ket{\Phi_{M\nu(j)}}  \nonumber \\
&=&(M - \sum_{k(j)} s_{jk} )\,  \ket{\Phi_{M\nu(j)}} ,
\end{eqnarray}
and has a total angular momentum $\hat{J}\ket{\Phi_{M\nu(j)}}=J
\ket{\Phi_{M\nu(j)}}$
\begin{eqnarray}
J&=&2Mj+J_{\nu(j)} ,
\end{eqnarray}
where the contribution from unpaired electrons is given by
\begin{eqnarray} \hspace*{-0.6cm}
 J_{\nu(j)}=\sum_{k(j)} |\nu_{jk}|  (j+\nu_{jk} \, k)+ J_{\sf in}, 
 \mbox{ with } J_{\sf in}=\sum_{j=1}^{N_{\sf in}} i_j .
\end{eqnarray}
Thus, one can classify the eigenstates of $H_{{\sf G}j}$ according to
their total angular momentum $\hat{J}$ and $S^z_j$.  

The analysis above allows us to label eigenstates according to a filling
fraction $\nu$ (related to $S^z_j$)  {\it and} the  angular momentum
$J$. The filling fraction in the RG problem is, by analogy to a QH
system on a  disk (Eq. \eqref{fillingfrac}), defined as follows:
\begin{eqnarray}
\nu=\frac{N -1}{L-1}=\frac{p}{q} \leq 1,
\end{eqnarray}
with $p$ and $q$ relative prime numbers. For fixed $\nu$, the latter
relation constrains allowed values of $N $ to be separated by integer
multiples of $p$ since $L$ is given by the integer $L=\frac{q}{p} (N -1)
+1$.

The model Hamiltonian of Eq. \eqref{modlppip} is exactly solvable, 
\cite{Romb, Ibanez} meaning that its {\it full} spectrum can be
determined with algebraic complexity.  In the present paper, because of
its relevance to QH physics, we are interested in a particular singular
limit of that model. We will consider the case where 
$\bar{g}=-2/(2M+N_{\nu(j)}-{\cal C}(j)-2)$, leading to a term in the
strongly-coupled Hamiltonian of Eq. (\ref{HLLL}), \cite{Pan}
\begin{eqnarray}
\hspace*{-0.7cm}
H_{{\sf G}j;m}&&=g \sum_{k(j),l(j)} \eta_{k} \eta_{l} \,
c^\dagger_{j+k} c^\dagger_{j-k} c^{\;}_{j-l} c^{\;}_{j+l} \nonumber \\
&&= g \ T_{j 1;m}^+ T_{j 1;m}^- ,
\label{HamiltG}
\end{eqnarray}
with $g = \lambda_{j} \overline{g}$. In each sector of fixed pair angular 
momentum $2j$ and for each pseudopotential index $m$,
 the  LLL QH Hamiltonian is
{\it identical} to the QH-RG Hamiltonian of Eq. (\ref{HamiltG}). We will
return to the investigation of the full QH problem  formed by sums of
individual QH-RG Hamiltonians (see Eq. (\ref{HLLL}) in Section
\ref{qha}). For the time being, we remark that the individual QH-RG
Hamiltonians corresponding to different values of $j$ generally do not
commute with one another; there are only four QH-RG  Hamiltonians
$H_{{\sf G}j} $ which are special in that they are diagonal; these
correspond to $j=\frac{1}{2}, 1, j_{\sf max}-\frac{1}{2}, j_{\sf max}$
with $ j_{\sf max}$ denoting the maximal possible value of $j$ 
(see Eq. \eqref{jvalues}). 

It is notable that contrary to more standard pairing problems,
especially those in which pairing may arise in mean-field treatments,
when a Haldane pseudopotential is used, the Hamiltonian of Eq.
(\ref{HLLL}) is {\it repulsive} i.e., $g>0$. Nonetheless,  as we
elaborate on below, in the decomposition into exactly solvable QH-RG 
Hamiltonians, we will find that each repulsive term associated with a
given $j$ (and $m$) in  Eq. (\ref{HLLL}), pairing is induced in the sense 
that pair fluctuations dominate correlations among electrons.

\subsection{Hilbert space analysis}
\label{Hilbert}

Given $N $ spinless fermions and $L$ orbitals, the dimension  of the Hilbert
state space ${\cal H}_L(N )$ is
\begin{eqnarray}
\dim {\cal H}_L(N )=\binom{L}{N }=\frac{L!}{N ! (L-N )!} . 
\end{eqnarray}
The set of allowed total angular momenta $J$ is given by
\begin{eqnarray}
{\cal J}_L(N)&=&\Big \{\frac{N (N -1)}{2},\frac{N (N -1)}{2}+1,
\nonumber \\
&&\hspace*{-1.5cm}\frac{N (N -1)}{2}+2, \cdots ,  N \left ( L - \frac{(N
+1)}{2}\right ) \Big \} , 
\end{eqnarray}
such that 
\begin{eqnarray}
\dim {\cal H}_L(N )=\sum_{J \in {\cal J}_L(N)} \dim {\cal H}_L(N ,J), 
\end{eqnarray}
where ${\cal H}_L(N ,J)$ is the Hilbert subspace with fixed total
angular momentum $J$. 

Given a fixed number of electrons $N $, of orbitals $L$, and angular
momentum $J$,  one can determine the dimension of the Hilbert space
${\cal H}_L(N ,J)$ as follows:  The dimension of ${\cal H}_L(N ,J)$ is
equal to the total number, $N_{\{m_i\}}$,  of  distinct partitions
$\{m_i\}$=$\{m_1,m_2,\cdots, m_{N }\}$, $\sum_{i=1}^{N }m_i=J$, of the
integer $J$,  and  can be determined  with the help of the following
generating function
\begin{eqnarray}
\label{fermipart}
\hspace*{-0.5cm}
{\cal{Z}}(x,z)=\prod_{r=0}^{m_{\sf max}} (1+ z x^r) =
\sum_{\bar{J}=0}^{\bar{m}_{\sf max}}  \sum_{\bar{N}=0}^{m_{\sf max}+1}
{\cal P}(\bar{N},\bar{J}) \, z^{\bar{N}}  x^{\bar{J}},
\end{eqnarray}
where $m_{\sf max}=L-1$ is the largest integer that
may appear in the partition $\{m_i\}$,  and $\bar{m}_{\sf max}=m_{\sf
max}(m_{\sf max}+1)/2=(L-1)L/2$. The dimension of ${\cal H}_L(N ,J)$ is
\begin{eqnarray}
\hspace*{-0.5cm}
\dim {\cal H}_L(N ,J)=N_{\{m_i\}}={\cal P}(N,J).
\end{eqnarray}
The number of partitions associated with the filling fraction of $\nu
=1$, see Eq. (\ref{fillingfrac}), constitutes a limiting non-vanishing
value, ${\cal P}(N, N (N -1)/2) =1$. [This single possible partition
corresponds to the arithmetic series, $\sum_{r=0}^{N -1}r = N  (N
-1)/2$.] 

We note that Eq. (\ref{fermipart}) corresponds to the grand
canonical partition function of a system of free spinless fermions with
equally spaced single particle energy levels similar to a harmonic
oscillator system, and trivially constrained by a cutoff $m_{\sf max}$. 
That is, in Eq. (\ref{fermipart}), $z$ may be regarded as the fugacity 
($e^{\beta \mu}$ with $\mu$ the chemical potential)  of these particles
and $x$ as the Boltzmann factor associated with the equally spaced
levels ($x= e^{- \beta \varepsilon}$ with a linear energy dispersion 
$\varepsilon_{r}= r \varepsilon$,  and inverse temperature $\beta$).
Such equally spaced levels are formally similar to those of the original
Landau level problem of non-interacting spinless fermions  in a magnetic
field (yet now sans a degeneracy of the single  particle states). In our
case, unlike that of standard non-interacting fermion problems, the
equally spaced levels may only be occupied up to a threshold value,
i.e., up to $r=m_{\sf max}$. 
This cutoff constraint is trivial and does not affect the Fermi function
occupancy of levels which we will shortly discuss below (formally, such
a cutoff may also be implemented by setting the energies of all
non-allowed levels to be positive and infinite for which the
corresponding Fermi function trivially vanishes as it must).

In the canonical ensemble  one has  to place $N $ fermions,
\begin{eqnarray}
N  = \sum_{r=0}^{m_{\sf max}} n_{r}
\label{ne}
\end{eqnarray}
 over $(2j+1) = m_{\sf max}+1$ levels such that the total ``energy''
\begin{eqnarray}
\label{jtot}
J = \sum_{r=0}^{m_{\sf max}} \varepsilon_r \ n_{r}
\end{eqnarray}
is fixed, with occupancies $n_{r}=0,1$.  The total number of states is
given by
\begin{eqnarray}
\label{NS}
{\cal P}(N,J) = e^ {\sf S},
\end{eqnarray}
where the entropy ${\sf S}$ is  defined by the corresponding entropy of
the Fermi-Dirac gas with a linear energy dispersion, and in  units such
that $k_{B}=1$.

It is clear that the number of partitions increases exponentially for
large system sizes. A quantitative approximation for this increase can
be obtained  in the grand canonical ensemble in the relevant
thermodynamic limit. Let us start defining the average number of
particles
\begin{eqnarray}
\hspace*{-0.5cm}
\bar{N} = z \frac{\partial \ln {\cal Z}}{\partial z} =
\sum_{r=0}^{m_{\sf max}} \langle n_{r} \rangle  ,  \nonumber \\
\mbox{ with } \langle n_{r} \rangle \equiv \langle n_{\varepsilon} \rangle =
\frac{1}{1+e^{\beta(\varepsilon-\mu)}}
\label{neav}
\end{eqnarray}
representing the mean occupation number,  and average ``energy''
\begin{eqnarray}
\hspace*{-0.5cm}
\bar{J} = - \frac{\partial \ln {\cal Z}}{\partial \beta} =
\sum_{r=0}^{m_{\sf max}}  \varepsilon_{r} \ \langle n_{r} \rangle  ,
\label{eav}
\end{eqnarray}

The equally spaced levels $\varepsilon_{r} = r \epsilon$ imply a
constant density of states (of size unity) in approximating the discrete
sums in Eqs. (\ref{neav}) and (\ref{eav}) by integrals from which it is
seen that average number of particles $\bar{N}$ and average ``energy''
are moments of  the Fermi function $\langle n_{\epsilon} \rangle$.
\begin{eqnarray}
\bar{N}& \approx &  \int_{0}^{m_{\sf max}} d\varepsilon \ \langle
 n_{\varepsilon} \rangle \ , \
\bar{J}  \approx  \int_{0}^{m_{\sf max}} d\varepsilon \ \varepsilon \,
\langle n_{\varepsilon} \rangle .
\label{NJ}
\end{eqnarray}
Further
corrections to the integral approximations above to the original sums
over discrete states may be obtained via the Euler-Maclaurin formula.

In Eq. (\ref{NJ}), $\mu$ and $\beta$ are Lagrange multipliers that fix
the  averages in Eqs. (\ref{ne}) and (\ref{jtot}). The integrals of Eqs.
(\ref{NJ}) are readily evaluated,
\begin{eqnarray}
\bar{N} & \approx &   m_{\sf max} + \frac{1}{\beta}  \ln \Big(
\frac{1+ e^{-\beta  \mu}}{1+ e^{\beta (m_{\sf max}- \mu)}} \Big),
\nonumber \\
\bar{J} & \approx & \frac{1}{2 \beta^{2}} \Big[ \beta^{2} m_{\sf
max}^{2} - 2 \beta m_{\sf max}  \ln (1+ e^{\beta(m_{\sf max}- \mu)}))
\nonumber \\
&-& 2 \, {\sf Li}_{2} (-e^{\beta(m_{\sf max}-\mu)}) + 2  \, {\sf
Li}_{2}(-e^{-\beta \mu})\Big],
\label{NJ_long}
\end{eqnarray}
where ${\sf Li}_{2}(z) = \sum_{a=1}^{\infty} \frac{z^{a}}{a^{2}}$ is the
polylogarithmic function of order two. Specializing to an incompressible
Laughlin fluid, if we set $N  = \bar{N}$ and $m_{\sf max} =
\frac{q}{p}(\bar{N}-1)$, we find, in this thermodynamic limit (whence
we  approximate $(\bar{N} -1) \sim \bar{N}$)) that, from the first of
Eqs. (\ref{NJ_long}),
\begin{eqnarray}
\label{fugacity_free_gas}
e^{\beta \mu} \approx \frac{1 - e^{\beta
\bar{N}}}{e^{(1-\frac{q}{p})\beta \bar{N}}-1}.
\end{eqnarray}
Formally, for the particular case of $\frac{q}{p}=2$, Eq.
(\ref{fugacity_free_gas}) further simplifies to $\mu = \bar{N}$.  Given
also $\bar{J}$, the combination of Eq. (\ref{fugacity_free_gas}) and the
second of  Eqs. (\ref{NJ_long}) provides both $\beta$ and $\mu$.

The entropy of the free Fermi system is the sum of the entropies
associated with that of the decoupled levels $\varepsilon$ (for which the
probabilities of the two possible states  (i.e., of having the state of
energy $\varepsilon$ being occupied or empty)  are $\langle
n_{\varepsilon} \rangle $ and $(1- \langle n_{\varepsilon} \rangle)$
respectively)  and is thus given (in the continuum integral
approximation to the original discrete $\varepsilon$ sums) by
\begin{eqnarray} \hspace*{-0.8cm}
{\sf S}= - \tr \rho \ln \rho \approx -  \int_{0}^{m_{\sf max}}
d\varepsilon \Big[ \langle  n_{\varepsilon} \rangle \ln\langle
n_{\varepsilon} \rangle \nonumber \\
+  (1-\langle n_{\varepsilon} \rangle) \ln(1-\langle n_{\varepsilon}
 \rangle) \Big],
 \label{entrop}
\end{eqnarray}
where $\rho$ is the density matrix.  Armed with the entropy of Eq.
(\ref{entrop}), we may next invoke Eq. (\ref{NS}) to compute  the number
of states (i.e., Hilbert space dimension). It is readily seen that the
entropy is extensive in $m_{\sf max}$ and thus the system  size $N $.
Unfortunately, an illuminating closed form expression is not attainable.

\subsection{Eigenspectrum}
\label{eigenspectrum}

The model Hamiltonian $H_{{\sf G}j}$ of Eq. \eqref{HamiltG} is exactly
solvable, meaning that  one can write down its full eigenspectrum with 
algebraic complexity.  
 The (unnormalized) eigenvectors of $H_{{\sf G}j}$ are the
states
\begin{eqnarray}
\ket{\Phi_{M\nu(j)}}=\prod_{\alpha=1}^M {\sf
S}^+_{j}(E_\alpha)\ket{\nu(j)} ,
\label{statepaired}
\end{eqnarray}
with
\begin{eqnarray}
{\sf S}^+_{j}(E_\alpha)=\sum_{k(j)}
\frac{\eta_{k}}{\eta_{k}^2-E_\alpha} \ c^\dagger_{j+k}
c^\dagger_{j-k},
\end{eqnarray}
and where the seniority eigenstates $\ket{\nu(j)}$ satisfy  the relation
$H_{{\sf G}j}\ket{\nu(j)}=0$. Note that the structure of these equations is the 
same for different pseudopotential indices $m$, and only depends on the 
general factorized form of the Hamiltonian. To avoid cumbersome notation, as we 
have done  in last
sections, we will often omit the pseudopotential rank index $m$.

The eigenvalue equation can then be written as
\begin{eqnarray}
H_{{\sf G}j}\ket{\Phi_{M\nu(j)}}&=&[H_{{\sf
G}j},\prod_{\alpha=1}^M {\sf S}^+_j(E_\alpha)] \ket{\nu(j)}
\nonumber \\
&=& {\cal E}_{M\nu(j)} \ket{\Phi_{M\nu(j)}},
\end{eqnarray}
with the commutator
\begin{eqnarray}
\label{commutaRG}
[H_{{\sf G}j},\prod_{\alpha=1}^M {\sf S}^+_{j}(E_\alpha)] &= & \\
&&\hspace*{-3.cm} -2g \sum_{\alpha=1}^M \mathbb{S}^+_{j\alpha} \ \Big (
\sum_{k(j)} \frac{\eta^2_{k}}{\eta^2_{k}-E_\alpha} S^z_{jk}+ 
\sum_{\beta (\neq \alpha)=1}^M \frac{E_\beta}{E_\beta-E_\alpha} \ \Big )
\nonumber
\end{eqnarray}
and
\begin{eqnarray}
\mathbb{S}^+_{j\alpha}= \left (\prod_{\gamma(\neq \alpha)=1}^M {\sf
S}^+_{j}(E_\gamma)  \right ) \ T^+_{j1} .
\end{eqnarray}

\begin{figure*}[htb]
\includegraphics[width=1.4\columnwidth]{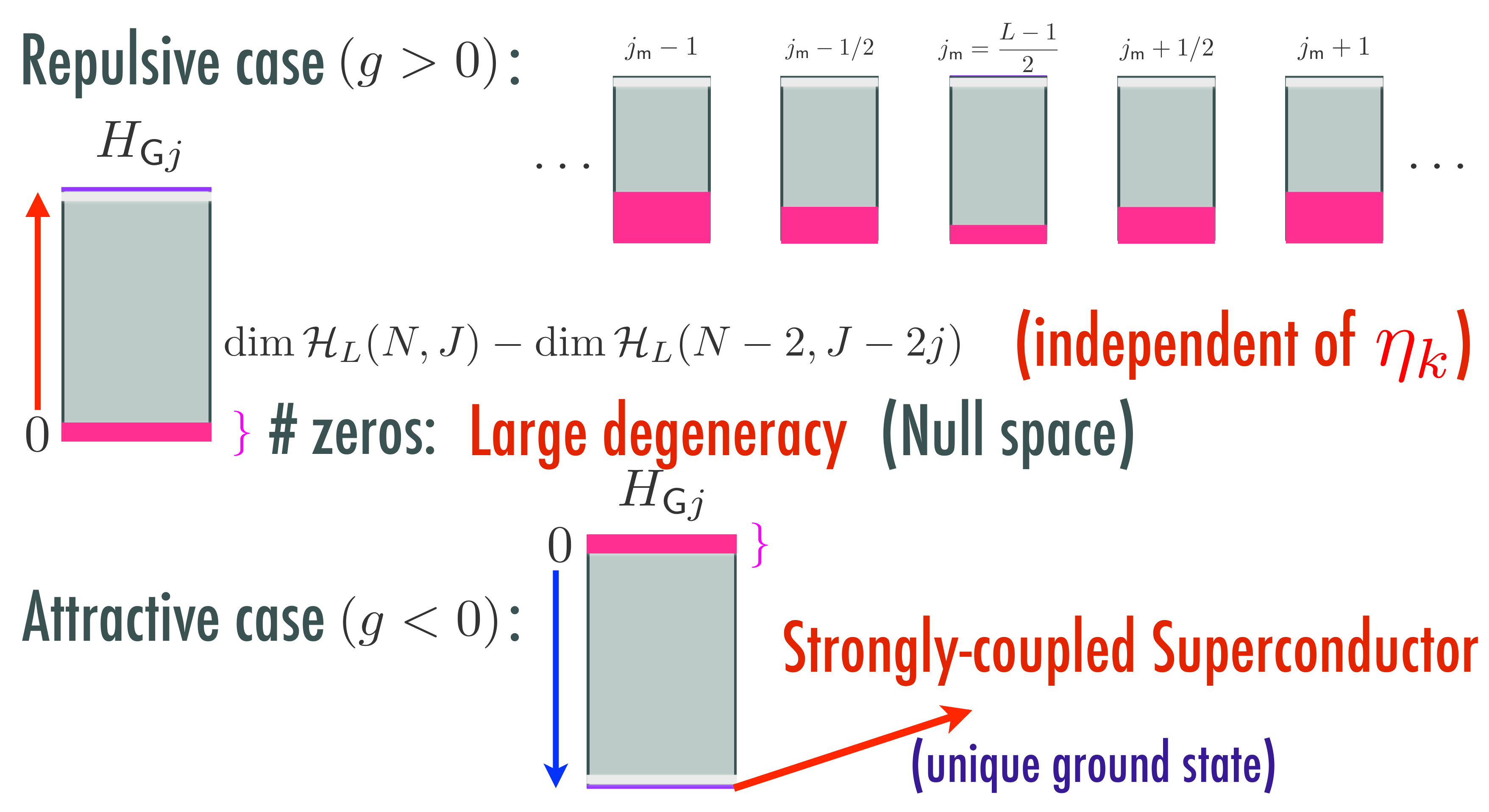}
 \caption{Eigenvalue spectrum of the repulsive and attractive QH-RG
 model,  which is the strong-coupling limit of the hyperbolic
 ($p_x+ip_y$) RG model.  }
 \label{Spectrum.fig}
\end{figure*}

There are two distinct types of solutions:

$\bullet$ It is clear from the commutator, Eq. \eqref{commutaRG}, that zeroing the
quantity  in parentheses there are solutions with eigenvalue (see Fig.
\ref{Spectrum.fig})
\begin{eqnarray}
{\cal E}_{M\nu(j)}=0,
\end{eqnarray}
corresponding to the case where {\it all} the spectral parameters
$E_\alpha$ (also known as pairons) are {\it finite} (complex-valued, 
in general).   
The RG (Bethe) equations satisfied by those pairons
are of the form,
\begin{eqnarray}
\sum_{\beta (\neq \alpha)=1}^M  \frac{E_\beta}{E_\beta-E_\alpha}
-\sum_{k(j)}  s_{jk} \ \frac{\eta^2_{k}}{\eta^2_{k}-E_\alpha} =0
, \ \forall \alpha
\label{RGeqs0}
\end{eqnarray}
$\alpha \in [1,M]$,  which can be re-written as (when $E_\alpha \neq
0$)
\begin{eqnarray}
\sum_{k(j)}   \frac{s_{jk}}{\eta^2_{k}-E_\alpha} -\sum_{\beta (\neq
\alpha)=1}^M  \frac{1}{E_\beta-E_\alpha} -\frac{Q_j}{E_\alpha} =0,  \
\forall \alpha
\label{RGeqs}
\end{eqnarray}
with $Q_j=M-1- \sum_{k(j)} s_{jk} =M-1+(N_{\sf b}-{\cal C}(j))/2$.  If
there is one vanishing pairon, $E_\alpha=0$, then the following
condition needs to be satisfied
\begin{eqnarray}
\sum_{k(j)}   s_{jk}=M-1 .
\end{eqnarray}

$\bullet$ There is another class of solutions that corresponds to having  one
pairon $E_M \rightarrow \infty$, where
\begin{eqnarray}
{\sf S}^+_j(E_M)  \rightarrow - \frac{1}{E_M} \ T^+_{j1} ,
\end{eqnarray}
with the remaining pairons ($\alpha \in [1,M-1]$) being
finite-valued and satisfying the RG equations
\begin{eqnarray}
1+\sum_{\beta (\neq \alpha)=1}^{M-1}  \frac{E_\beta}{E_\beta-E_\alpha}
-\sum_{k(j)}  s_{jk} \ \frac{\eta^2_{k}}{\eta^2_{k}-E_\alpha} =0 .
\label{RGeqsM}
\end{eqnarray}
For this class of solutions the corresponding eigenvalues of $H_{{\sf
G}j}$ are {\it positive(negative)} (see Fig. \ref{Spectrum.fig})
\begin{eqnarray}
{\cal E}_{M\nu(j)}=2g \left (\sum_{k(j)}  s_{jk} \, \eta^2_{k}  -
\sum_{\alpha=1}^{M-1} E_\alpha \right ) ,
\label{nonzeroeig}
\end{eqnarray}
which simply results from the fact that $H_{{\sf G}j}$ is a
positive(negative) semi-definite operator  when $g > 0$  $(g < 0)$.

Care has to be taken with the different seniority subspaces entering in
Eqs. \eqref{RGeqs0}  and \eqref{nonzeroeig} through the eigenvalues
$s{_{jk}}$ of the  $S^z_{jk}$ operator, Eq. \eqref{Szjk}. The total
number of unpaired electrons  $N_{\nu(j)}$ should have a total angular
momentum $J_{\nu(j)}= j N_{\sf b}+J_{\sf in}$,  and they should not
couple in pairs to angular momentum $2j$. Therefore, the  seniority
configuration of  $N_{\sf b}$ electrons $[j+k_{1}, j+k_{2},\cdots,
j+k_{N_{\sf b}}]$  must fulfill the condition
\begin{equation}
\sum^{N_{\sf b}}_{i=1} k_i = 0.
\label{full}
\end{equation}

We note here that for any seniority configuration  $[j+k_{1},\cdots,
j+k_{2}, j+k_{N_{\sf b}}]$ satisfying the condition  (\ref{full}) there
is another seniority configuration  $[j-k_1, j-k_2, \cdots, j-k_{N_{\sf
b}}]$ blocking the same pair states  ${\bf k}=[j+k,j-k]$, satisfying
(\ref{full}), and with same set of parameters $s_{jk}$. These two
solutions have the same energy Eq. \eqref{nonzeroeig}.  Hence, any
eigenvalue with one infinite pairon, $M-1$ finite pairons, and non-zero
seniority  is at least doubly degenerate.

One can analytically determine the largest contribution to the first
term in Eq. \eqref{nonzeroeig} for the non-zero eigenvalues
\begin{eqnarray}
\sum_{k(j)} \eta^2_{k},
\end{eqnarray}
corresponding to $s_{jk} =1/2$, for all values of $k$. For  the disk
geometry, for instance, it is given by
\begin{eqnarray}
\sum_{k(j)} \eta^2_{k}= \frac{1}{2^{2j-2} j} \sum_{k=1/2}^j  k^2
\binom{2 j}{j+k}.
\end{eqnarray}
The sum can be easily shown to be given by
\begin{eqnarray}
\sum_{k=1/2}^j  k^2 \binom{2 j}{j+k}=j \, 2^{2j-2} ,
\end{eqnarray}
implying that the largest contribution is
\begin{eqnarray}
\label{sumpi}
\sum_{k(j)} \eta^2_{k}= 1,
\end{eqnarray}
a trivial constant value independent of $L$ and $N$. This normalization
makes explicit earlier considerations which led to Eq. (\ref{etanorm}).

Inspection of Eqs. \eqref{RGeqs0} or \eqref{RGeqsM} tells us that the
set of spectral parameters $\{ E_\alpha \}$ is identical to the set  $\{
E_\alpha^* \}$, meaning that the pairons are either real-valued or if a
pairon, e.g., $E_1$, is complex then there exists another pairon
solution that is its complex conjugate, i.e.,  $E_1^*$. Notice that the
RG equations, and consequently the spectral parameters, do not depend on
the coupling strength $g$.

Therefore, all non-zero energy eigenstates are associated with  spectral
parameters which are all finite-valued except one,  identified  with
$E_M$, which becomes infinite.  Because of the latter, the total number 
of positive(negative) energy
eigenstates is given by the number of partitions ${\cal P}(N -2, J-2j)$, which implies that the
total number of zero energy eigenstates is $N_{\sf z}={\cal
P}(N,J)-{\cal P}(N -2, J-2j)$. Table \ref{table:dimensions} displays
some characteristic values  of various dimensions for systems up to
$N$=10 electrons. 

\begin{table}[hbt]
\begin{tabular}{ c|c|c|c|c|c}
\hline\hline
$N $  & $m_{\sf max}$   &  $J_{\sf m}$ & $N_{\{m_i\}}$ & $N_{\sf z}$ &
$\ell_{\frac{N }{2}}$ \\ \hline
2 & 3  & 3 & 2 & 1 & 1\\
4 & 9  & 18 & 18 & 13 & 5\\
6 & 15  & 45 & 338 & 252 & 28\\
8 & 21  & 84 & 8512 & 6375 & 165\\
10 & 27  & 135 & 246448 & 184717 & 1001 \\
\end{tabular}
\caption{Dimension of the Hilbert space, $N_{\{m_i\}}$, and number of
zero energy eigenstates, $N_{\sf z}$,  $\ell_M=
\binom{{\cal C}(j)}{M}-\binom{{\cal C}(j)}{M-1}$, for $\frac{q}{p}=3$ and
$2j_{\sf m}=m_{\sf max}$. }
\label{table:dimensions}
\end{table}

The nature of the ground state of $H_{{\sf G}j}$ depends on the sign of
$g$. In the repulsive ($g>0$) case, the ground state is, in general,
highly degenerate and its energy is zero regardless of the system size.
On the contrary, in the attractive ($g<0$) case the ground state energy
is negative, non-degenerate, and grows in magnitude with system size
according to Eq. \eqref{nonzeroeig} (see Fig. \ref{Spectrum.fig}).

%

\subsection{Symmetry properties of the RG equations}
\label{sym_prop}

In this section we are interested in analyzing the consequences of
having vanishing spectral parameters, i.e., a set of  pairons with
$E_\alpha=0$. This analysis unveils a {\it symmetry relation} of the RG
equations that connects eigenstates with different filling fractions
$\nu$. 

Consider an $M$ pair state of the form ($\tilde{M}=M-N_0$)
\begin{eqnarray}
\hspace*{-0.7cm}
\ket{\Phi_{M \, \nu(j)}} =({\sf S}_j^+(0))^{N_0}\ket{\Phi_{\tilde{M}
\nu(j)}}= (T^+_{j 0})^{N_0} \ket{\Phi_{\tilde{M} \nu(j)}} ,
\end{eqnarray}
where we assume that $N_0$ pairons vanish, and $\ket{\Phi_{\tilde{M}
\nu(j)}}$ is an eigenstate of $H_{{\sf G} j}$. What are the conditions
necessary for $\ket{\Phi_{M \, \nu(j)}}$ to be an eigenstate of $H_{{\sf
G} j}$? To address this question one needs to evaluate the commutator
\begin{eqnarray}
\hspace*{-0.7cm}
[H_{{\sf G}j},(T^+_{j 0})^{N_0}]&=&-g {N_0} (T^+_{j 0})^{{N_0}-1} T^+_{j
1}(2 T^z_{j0}+{N_0}-1) ,
\end{eqnarray}
with $T^z_{j0}=S^z_{j}$. Since $\ket{\Phi_{\tilde{M} \, \nu(j)}}$ is
also an eigenstate of $S^z_{j}$, the vanishing of this commutator would
indicate that the states $\ket{\Phi_{\tilde{M} \, \nu(j)}}$ and
$\ket{\Phi_{M\nu(j)}}$ are degenerate, i.e., share the same eigenvalue
although they correspond to different filling fractions.

It follows that if the number of vanishing spectral parameters satisfies
\begin{eqnarray}
{N_0}=2(M-\sum_{k(j)}s_{jk})-1=1+2Q_j,
\end{eqnarray}
then the states $\ket{\Phi_{M\nu(j)}}$ and $\ket{\Phi_{\tilde{M} \,
\nu(j)}}$ are degenerate. Moreover, no pairons $E_\alpha$ converge  to
zero for $2(\sum_{k(j)}s_{jk}-M)+1 \geq 0$. Note that the filling
fractions corresponding  to these two states are
\begin{eqnarray}
\nu^M = \frac{N (N -1)}{2J} \ , \
\nu^{\tilde{M}}=  \frac{\tilde{N}(\tilde{N}-1)}{2\tilde{J}},
\end{eqnarray}
with $\tilde{N}=N -2{N_0}$ and $\tilde{J}=J-2{N_0}j$.

This symmetry relation, which is independent of the sign of the coupling
$g$, has interesting and important consequences. (For a related 
discussion in the context of the $(p_x+ip_y)$ superconductor  
see Ref. \onlinecite{Romb}.) Consider the two special cases:
\begin{enumerate}
\item
\underline{Symmetric case}: In this limiting case $M=\tilde{M}$ 
(i.e., there are no zero-valued pairons, $N_{0}=0$)
\begin{eqnarray}
M=\frac{1+{\cal C}(j)-N_{\sf b}}{2} \Rightarrow Q_j=-\frac{1}{2} .
\end{eqnarray}
For attractive interactions ($g<0$), this limiting case is associated 
with a non-trivial quantum critical point signaling a topological 
zero-temperature phase transition in the thermodynamic limit. \cite{Romb} 
\item
\underline{Asymmetric case}: ${N_0}=M-1$ (all but one zero-valued pairons)
\begin{eqnarray}
M={\cal C}(j)-N_{\sf b}  \Rightarrow Q_j=\frac{M-2}{2} .
\end{eqnarray}
\end{enumerate}

To get an understanding of the meaning of these particular relations,
consider the case of seniority zero eigenstates, i.e.,  $N_{\nu(j)}=0$,
$N =2M$, leading to $\nu^M=(2M-1)/(2j_{\sf m})$. Then, the symmetric
case corresponds to $\nu^M=\nu^{\tilde{M}}={\cal C}(j_{\sf m})/(2 j_{\sf
m}) (\stackrel{}{\rightarrow} 1/2)$, while the asymmetric case
corresponds to $\nu^M=(2 {\cal C}(j_{\sf m})-1)/(2j_{\sf m}) 
(\rightarrow 1)$, and $\nu^{\tilde{M}}=1/(2j_{\sf m}) (\rightarrow 0)$.
The  values displayed in parentheses correspond to the  large $j$ limit,
with $j$'s  such that ${\cal C}(j)=[j+\frac{1}{2}]$.

\section{Ground States of the full pseudopotential problem}
\label{full_QH}

In this section, we survey some known results pertaining to the zero
energy ground states of Haldane pseudopotentials, and rigorously
generalize some of these results using our second quantized
formulation.  In the earlier sections we analyzed the problem for fixed
$m$ and $j$ RG type  Hamiltonians $H_{{\sf G}j;m}$. We now turn to the
full problem formed by the sum of these Hamiltonians over all $m$ and
$j$ (Eq. (\ref{HLLL})), and make use of Eq. (\ref{HamiltG}) to write a
generic rotationally symmetric Hamiltonian in the LLL as a sum of
QH-RG  Hamiltonians, 
\begin{eqnarray}
\label{HLLLG}
\widehat{H}_{\sf QH} =\sum_{m} \sum_{0<j<L-1} H_{{\sf G}j;m}. 
\end{eqnarray}
As we have shown in this work, for the usual pseudopotential
expansion, each term in the above sum is indeed of the RG form.
In the following, we will, however, also have opportunity to consider
generalizations where the $H_{{\sf G}j;m}$ are RG-terms with $\eta_k$'s
{\em not} necessarily corresponding to a Haldane pseudopotential.


For concreteness, in what follows, we first focus on the lowest ($m=1$)
pseudopotential.    The structure of many of the following considerations
is identical for all $m$.  
Generally, we will be interested in the case where the sum over $m$ in \Eq{HLLLG}
is finite.
The number of zero-energy states of Eq.
(\ref{HLLLG}) depends on how many terms with different $m$ and $j$ are included. 
We elaborate on this in  Section \ref{nullspace}.  We then
illustrate (Section \ref{inward})  how notions of ``inward squeezing"
can  be generalized to states that are defined through a Hamiltonian, 
rather than an analytic clustering property.  In Section \ref{RG0}, we
explain how the basis associated with the QH-RG Hamiltonians can be used
as a new basis to expand Laughlin states. Facts concerning the
conventional Slater determinant decomposition of Laughlin states are
reviewed and expanded on in Section \ref{slatersec}. We explicitly note
a cutoff value (in particle number) beyond which some ``admissible'' Slater determinant states have a
vanishing amplitude for the $\nu =1/3$ Laughlin state, underscore the
relevance of maximally paired configurations (central to our RG
approach), and further explicitly relate the squeezed state
formulation, on which we present some rigorous results in Section \ref{inward}, to 
``admissible'' (in a Young tableau sense\cite{Francesco}) Slater determinant states. 

\subsection{Null space and frustration-free properties of $H_{V_1}$}
\label{nullspace}

{}From previous sections we conclude that the QH Hamiltonian can be
written as a direct sum of hyperbolic QH-RG Hamiltonians 
\begin{eqnarray}
\hspace*{-0.7cm}
H_{V_1}&=&\!\!\!\sum_{0< j < L-1} H_{{\sf G}j} \ ,
\label{v1sum}
\end{eqnarray}
with, in general,
\begin{eqnarray}
[H_{{\sf G}j},H_{{\sf G}\bar{\jmath}}]\neq 0 \ \ \ (j\neq \bar{\jmath}).
\label{v1commute}
\end{eqnarray}
In this equation, we have fixed the pseudopotential index $m=1$ and
simply denote $H_{{\sf G}j;m=1} = H_{{\sf G}j}$.    The gauge symmetry
of  Eq. \eqref{gaugesym} displayed by each $H_{{\sf G}j}$ is no longer
a   symmetry of the QH Hamiltonian  $H_{V_1}$,  thus seniority is not
conserved. Nonetheless, since Laughlin states  are exact ground states,
as we discussed in detail above, the Hamiltonian is still  quasi-exactly
solvable, at least for $\nu=1/{3}$ (and $\nu<1/3$). By
this we mean that the ground state(s) can be determined exactly, and is(are)
related to the integrable structure that we  exposed above, but no such
characterization is known for the finite energy  excited states.

We are interested in understanding the properties of the null space
${\sf Ker}(H_{V_1})$.  
In the following sections, we wish to establish a series of exact analytic
properties that emerge from our second quantization analysis.  Let us
start with the following known result, which we paraphrase as follows:\cite{Haldane,Trugman}  ``Given
$L, N $, the  Hamiltonian  $H_{V_1}$ displays zero energy ground states
$\ket{\Psi_\nu^J}$, i.e. $H_{V_1} \ket{\Psi_\nu^J}=0$,  whenever $L \geq
3N  -2$, or equivalently, $0 \leq \nu=\frac{p}{q}\leq\frac{1}{3}$.  The
zero energy state is unique when $\nu=\frac{1}{3}$, it is in the sector 
$J=J_{\sf m}$, and is the Laughlin state $\ket{\Psi^{J_{\sf
m}}_{\frac{1}{3}}}$''.  Armed with this result, one can state a
remarkable property of the  null space ${\sf Ker}(H_{V_1})$: 
``$H_{V_1}$ is a frustration-free Hamiltonian for $0 \leq \nu
\leq\frac{1}{3}$''. This means that ${\sf Ker}(H_{V_1})$ is the  {\it
common} null space of {\it all} the null spaces ${\sf Ker}(H_{{\sf
G}j})$. 

The proof goes as follows: The states $\ket{\Psi^J_\nu}$ are zero energy
ground states of $H_{V_1}$, which is a direct sum of positive
semi-definite operators $H_{{\sf G}j}$. Therefore,
\begin{eqnarray}
H_{{\sf G}j} \ket{\Psi_\nu^J} = 0 , \ \mbox{ for all } j , \ j_{\sf min}
 \leq  j \leq j_{\sf max}
\end{eqnarray}
i.e., $\ket{\Psi_\nu^J}$ are zero energy ground states of each RG
Hamiltonian $H_{{\sf G}j}$. Moreover, $0=\langle\Psi^J_\nu|H_{{\sf
G}j}|\Psi^J_\nu\rangle =g^2 \|T^-_{j1}|\Psi^J_\nu\rangle\|^2$ implies
$T^-_{j1}|\Psi^J_\nu\rangle=0$ for all $j$, and filling fraction 
$\nu \le 1/3$. 


The results above generalize to 
Hamiltonians of the form 
$\widehat{H}_{\sf QH}=\sum_{0\leq m\leq
{\sf M}} g_m H_{V_m}$, where the  $g_m$ are positive for
$(-1)^m=(-1)^{\sf M}$ and otherwise $0$. Then, the  zero modes of this
Hamiltonian are simultaneously annihilated by  each operator
$H_{V_{m,j}}$ defined in Section \ref{V_n}. This condition  is
satisfiable for $\nu\leq 1/({\sf M}+2)$, and right at filling factor 
$\nu=1/({\sf M}+2)$ is  satisfied uniquely by the Laughlin state
$|\Psi_{\nu}^{J_{\sf m}}\rangle$.\cite{Haldane}  

We emphasize that presently, to the best of our knowledge, this
frustration-free  property cannot be derived from algebraic properties
of the operators  $H_{{\sf G}j}$ alone. Instead, the proof relies
crucially on establishing the existence of ${\sf Ker}(H_{V_1})$ using
first quantized language. It is worth noting that in going back to
first  quantized language, the problem is embedded in a larger Hilbert
space that also contains  degrees of freedom associated with dynamical
momenta. It is only through an intricate  interplay between guiding
center degrees of freedom and dynamical momenta that the  known
analytical properties of Laughlin wave functions
result.\cite{Haldane2011}  This could hardly have been guessed from the
second quantized Hamiltonian  Eq. \eqref{HV1}  alone, which describes
only the guiding center variables. It is only for  the right choice of
orbitals $\phi_r(z)$, defined by the kinetic energy Hamiltonian  $H_K$
and not by the second quantized pseudopotential Eq. \eqref{HV1}, that
the  zero energy ground state of the problem can be characterized by
simple analytic properties.

Note that the QH Hamiltonian {\it differs} in a crucial
way from more standard frustration-free
Hamiltonian studied in the literature. \cite{frustration_free}
 In those cases the null space of the underlying
local operators can be trivially characterized. It is then only the
existence of a {\em common} null space, ${\sf Ker}(H)$, which is
non-trivial. In the QH case,  each QH-RG Hamiltonian $H_{{\sf G}j}$ is
not strictly  local but decays exponentially and, in addition, displays
a different number of pair operators for different values of $j$. The
null space of  each $H_{{\sf G}j}$ can be exactly  determined, but this is
already a non-trivial problem since it requires a  Bethe ansatz instead
of a semi-simple Lie-algebraic solution.  

There are four QH-RG  Hamiltonians $H_{{\sf G}j} $ that are special
since  they are diagonal operators in the Fock basis, i.e., they commute
among themselves,  and  correspond to $j=\frac{1}{2}, 1,  j_{\sf
max}-\frac{1}{2}, j_{\sf max}$ (see Eq. \eqref{jvalues}). Consider an expansion of a zero
energy  ground states of $H_{V_1}$, $\ket{\Psi_\nu^J}$, in a normalized
Slater determinant (Fock) basis ($n_r=0,1$) 
\begin{eqnarray} \hspace*{-1cm}
\{ \ket{\{n\}}\}&=&\{\ket{n_{0}, n_1, \cdots, n_r, \cdots,
n_{L-1}}\}\nonumber \\ 
&=&\Big
\{ \frac{1}{\sqrt{N!}} \prod_{r=0}^{L-1}  (c^\dagger_r)^{n_r}\ket{0}\Big
\}, 
\label{SlaterDetFock}
\end{eqnarray}
with $\sum_{r=0}^{L-1} n_r=N $, and  $\sum_{r=0}^{L-1} r \, n_r=J$.
Then,  the following result follows: ``All zero energy states have zero
coefficients for the  basis states with $(n_0=1, n_1=1)$, $(n_0=1,
n_2=1)$, $(n_{L-3}=1, n_{L-1}=1)$, and  $(n_{L-2}=1, n_{L-1}=1)$,  in a
Slater determinant expansion''.  We note that this result is in
agreement with the principle of   ``inward squeezing''.
\cite{Bernevig1,Bernevig2}

The proof of this assertion is straightforward:  Assume that
$\ket{\Psi_\nu^J}$ has Slater determinant basis elements  with, e.g.,
$(n_0=1, n_1=1)$. Then, $H_{{\sf G}j=\frac{1}{2}} \ket{\Psi_\nu^J} \neq
0$,  since $H_{{\sf G}j=\frac{1}{2}}=g \eta^2_{\frac{1}{2}} \, n_1 n_0$,
which contradicts  the frustration-free condition of $H_{V_1}$. We can
apply the same argument  for the other three cases where the QH-RG
Hamiltonians correspond to  $j=1, j_{\sf max}-\frac{1}{2}, j_{\sf
max}$. 

It turns out that the last argument can be considerably generalized and 
applied to a large class of Hamiltonians, as we show in the following
section.

\subsection{Characterization of the ``incompressible filling factor'' 
and ``inward squeezing'' through the second quantized pseudopotentials}
\label{inward}

Due to the (in general) non-commutativity of the operators $H_{{\sf
G}j}$, the characterization of frustration free ground states of the
full Hamiltonian $\widehat{H}_{\sf QH}$ is a task that goes beyond the
analysis of Section \ref{strong}, where the eigenstates of the
individual operators  $H_{{\sf G}j}$ have been systematically studied.
For Haldane-pseudopotentials, the problem has been well-studied in first
quantization where, e.g., for $H_{V_1}$, the solutions are just the $\nu=1/3$
Laughlin state and its quasi-hole excitations at
$\nu<1/3$.\cite{Haldane,Trugman} There are no zero energy, hence
frustration free, ground states at $\nu>1/3$. Our goal here is to
understand  such properties as much as possible in terms of the  second
quantized operators $H_{V_{m,j}}$ discussed at length in previous
sections. From their second quantized form, it might not seem obvious
that these operators have any common zero energy states at all for some
appropriate range of $m$ and $j$, and for given $\nu$ and arbitrary
system size. Here we primarily want to understand within second quantized
language why, for instance,  for the $V_1$ pseudopotential, the
``incompressible'' filling factor $\nu=1/3$ is special. By ``special'',
we allude to the fact that there can be no common zero energy state for the
operators $H_{{\sf G}j;m=1}$ at filling factor $>1/3$. The analogous question
can be asked for the  parent Hamiltonian of the $\nu=1/{\sf q}$ Laughlin
state.  We emphasize, however, that our results in this section will
establish rigorous bounds for the (non)existence of zero modes for a
large class of Hamiltonians.  The second-quantized Haldane
pseudopotentials are merely special cases that satisfy these bounds. The
same is true for the solvable Hamiltonians of Ref.
\onlinecite{Nakamura}.

Moreover, the questions asked here will naturally lead us to  rigorously
prove a {\it squeezing
principle}\cite{haldanerezayi94,Bernevig1,Bernevig2} for the zero modes 
of a general class of model Hamiltonians. We begin with some general
notions related to squeezing. The reader unfamiliar with  this concept
will find a more detailed review in Section \ref{slatersec}.  We expand a
given state $|\psi\rangle$ into occupancy eigenstates
\begin{equation}\label{psiexpansion}
  |\psi\rangle = \sum_{\{n\}}C_{\{n\}} |\{n\}\rangle\,.
\end{equation}
where $ |\{n\}\rangle$ denotes an occupation number eigenstate
$|n_0,\dotsc, n_{L-1}\rangle$ as in Eq. \eqref{SlaterDetFock}. 
We call a state  $|\{n\}\rangle$ with $C_{\{n\}}\neq 0$ $\psi$-{\em
expandable} if there is a state $|\{n'\}\rangle$
with $C_{\{n'\}}\neq 0$ such that $|\{n\}\rangle$ and $|\{n'\}\rangle$
are related as follows:
\begin{eqnarray}
|\{n'\}\rangle&=& |n_0,\dotsc , n_{j_1-k}+1, \dotsc , n_{j_1}-1, 
\dotsc , n_{j_2}-1 , \nonumber \\
&&\hspace*{0.5cm} \dotsc , n_{j_2+k}+1,  \dotsc , n_{L-1}\rangle ,
\end{eqnarray}
where $j_1\leq j_2$, and $k>0$. That is, 
\begin{eqnarray}
\langle \{n'\}|c^\dagger_{j_1-k}
c_{j_2+k} ^\dagger c^{\;}_{j_2}c^{\;}_{j_1} |\{n\}\rangle\neq 0 .
\end{eqnarray}
We will call a state
$|\{n\}\rangle$ with  $C_{\{n\}}\neq0$ non-$\psi$-{expandable}  or just non-expandable if
$|\{n\}\rangle$ is not $\psi$-{expandable}. Further, we will say that
$|\{n\}\rangle$ satisfies the {\it generalized $r$-Pauli principle} if there
is no more than one particle in any $r$-consecutive orbitals.

Next we define the general class of operators to which our results will
apply. We focus on fermions for simplicity, but it should be clear that
analogous  results can be obtained for bosons.

Consider the operators $T_{j1;m}^-=\sum_{k(j)} \eta_{k}(j,m) c_{j-k}c_{j+k}$
as in Eq. \eqref{Tj}, where 
$1\leq m\leq {\sf M}$ and
we have restored the dependence on $j$ and
$m$ on the right hand side, subject to the constraint that $m$ and $\sf
M$ are both odd. 
We will say that the family of operators $T_{j1;m}^-$
has ``the independence property'' if for any  $j$ and for
$\ell=\min(({\sf M}+1)/2,[ j+\frac 12],[ L-\frac 12
-j])$, the $m$ distinct $\ell$-tuples
$(\eta_1(j,m),\dotsc,\eta_\ell(j,m))$ have a linear span of dimension
$\ell$ if  $j$ is integer, and similarly the $m$ $\ell$-tuples
$(\eta_{1/2}(j,m),\dotsc,\eta_{\ell-1/2}(j,m))$ if $j$ is half
odd-integer.

It is easy to see that in particular, the $T_{j1;m}^-$ of the
Haldane-pseudopotentials  $H_{V_m}$ have the independence property for
any $\sf M$, simply by appealing to the polynomial structure of the
corresponding coefficients $\eta_k(j,m)$ identified in Section
\ref{V_n}.

Our results are expressed by the following theorem and simple
corollaries:

{\bf Theorem:} 
Let the operators $T_{j1;m}^-$, $m=1,3,\dotsc ,{\sf M}$ satisfy the
independence property, and let $|\psi\rangle$ be annihilated by all
$T_{j1;m}^-$, $m=1,3,\dotsc , {\sf M}$, $j=1/2,\dotsc, L-3/2$. Then any
non-expandable basis state $|\{n\}\rangle$ in the expansion of
$|\psi\rangle$ satisfies the ${\sf M}+2$-Pauli principle.

{\it Proof}: For simplicity, we first consider the case ${\sf M}$=1.
Note that the independence property then reduces to $\eta_1(j,m=1)\!\neq
\! 0$ ($\eta_{\frac 12}(j,m=1)\! \neq \! 0$) for $j$ integer (half
odd-integer).

We will prove the statement by contradiction. Suppose $|\{n\}\rangle$ is
non-expandable and does not satisfy the $3$-Pauli principle. Then
$\{n\}$ contains a string $11$ or a string $101$. Consider the former
case. Then we have $|\{n\}\rangle=c^\dagger_{j+\frac
12}c^\dagger_{j-\frac 12}|\{\tilde n\}\rangle$ for some $j$, where
$|\{\tilde n\}\rangle$ has two particles less than $|\{n\}\rangle$, and
$\tilde n_{j\pm\frac 12}=0$. Further, for $k>\frac 12$, all states
$c^\dagger_{j+k}c^\dagger_{j-k}|\{\tilde n\}\rangle$ have zero
coefficient in $|\psi\rangle$, or else $|\{n\}\rangle$ would be
$\psi$-expandable. We thus have $\langle \{\tilde n\}|
T_{j1;1}^-|\psi\rangle= \eta_{\frac 12}(j,1)C_{\{n\}}\neq 0$. This
contradicts $T_{j1;1}^-|\psi\rangle=0$. Ruling out strings $101$ works
just the same.

To generalize to the case of arbitrary odd $\sf M$, we have to rule out
strings of the form $10^{s}1$, with $0^{s}$ representing a string of $s$
zeros, where $s=0,\dotsc, {\sf M}$. For given $j$, form the new linear
combination of operators $\widetilde T_{j1}^-=\sum_{m} a_m
T_{j1;m}^-=\sum_{k(j)} \tilde \eta_{k} c_{j-k}c_{j+k}$, which still
satisfies $\widetilde T_{j1}^-|\psi\rangle=0$. Consider integer $j$ and
odd $s$.  The  independence property is then exactly what guarantees
that we can always choose the $a_m$ such that $\tilde \eta_{k}=0$ for
$k=1,\dots ,(s-1)/2$ and  $\tilde \eta_{(s+1)/2}=1$. Similarly for
half-odd-integer $j$ and even $s$, where we can choose $\tilde
\eta_{k}=0$ for $k=1/2,\dots, (s-1)/2$ and  $\tilde \eta_{(s+1)/2}=1$.
The operators $\widetilde T_{j1}^-$ thus have a ``hollow core'',  and
allow one to contradict the assumption that a non-expandable
$|\{n\}\rangle$ has the pattern $10^{s}1$ just as we did in the case
${\sf M}=1$ above. This concludes the proof of the  Theorem.

For a general state $|\psi\rangle$, we now define its filling factor in
an $L$-independent manner as $\nu=(N-1)/n_{\sf max}(\psi)$, where
$n_{\sf max}(\psi)$ is the highest orbital index in $|\psi\rangle$ that
has non-zero probability of being occupied in $|\psi\rangle$. Our main
result is then the following:

{\bf Corollary 1:} 
Let the operators $T_{j1;m}^-$, $m=1,3\dots, {\sf M}$ be defined as in
the Theorem above. Then a state $|\psi\rangle$ annihilated by all
$T_{j1;m}^-$ has a filling factor $\nu\leq 1/({\sf M}+2)$.

{\it Proof}: Because of the finite dimensionality of the Hilbert space,
we can always find a non-$\psi$-expandable basis state $|\{n\}\rangle$.
The latter satisfies the ${\sf M}+2$-Pauli principle. The densest basis
state satisfying this generalized Pauli principle is clearly $10^{{\sf
M}+1}10^{{\sf M}+1}1\dotsc0^{{\sf M}+1}1$, which has filling factor
equal to $1/({\sf M}+2)$. This necessitates that $|\psi\rangle$ has a
filling factor less than or equal to that value.

The following Corollary establishes a notion of squeezing for any zero
mode of {\em any} Hamiltonian  $H=\sum_{m,j}  {T^+_{j1;m}}T_{j1;m}^-$
with operators satisfying the assumptions of the Theorem. This includes
Hamiltonians beyond the realm of pseudopotentials, such as those
considered in Ref. \onlinecite{Nakamura}. By ``squeezing'', we mean the
operations facilitated by the operators $c^\dagger_{j_1+k} c_{j_2-k}
^\dagger c^{\;}_{j_2}c^{\;}_{j_1}$, $j_1<j_2$,  and $k>0$, i.e., in
essence the inverse of the operation defining an expandable state above.
A state $|\{n\}\rangle$ can be ``squeezed'' from a basis state
$|\{n'\}\rangle$ if it can be obtained from $|\{n'\}\rangle$ by repeated
application of squeezing operations.

{\bf Corollary 2:}
Let $T_{j1;m}^-$, $m=1,3,\dots ,{\sf M}$ and $|\psi\rangle$ be defined
as in the Theorem. Then any basis state $|\{n\}\rangle$ having non-zero
coefficient in the expansion \eqref{psiexpansion} can be squeezed from a
basis state $|\{n'\}\rangle$ (not necessarily always the same) that
satisfies the ${\sf M}+2$-Pauli principle and that also has non-zero
coefficient.

{\it Proof}: A finite number of applications of operators of the form
$c^\dagger_{j_1-k} c_{j_2+k} ^\dagger c^{\;}_{j_2}c^{\;}_{j_1}$,
$j_1<j_2$, and $k>0$, on $|\{n\}\rangle$ must lead to a non-expandable
basis state $|\{n'\}\rangle$ (with non-zero coefficient, by definition),
due to finite dimensionality of the Hilbert space. Then $|\{n'\}\rangle$
must satisfy the ${\sf M}+2$-Pauli principle by the Theorem, and 
$|\{n\}\rangle$ can be squeezed from  $|\{n'\}\rangle$.

We note that the observation made in Section \ref{nullspace}, concerning
zero amplitude for all states of the form $|11\dotsc\rangle$,
$|101\dotsc\rangle$ in zero modes of $V_1$ is a special case of this
Corollary. It is clear that such states could not be squeezed from
states satisfying the $3$-Pauli principle. The Corollary more generally
implies the fact that many more Slater determinants have vanishing
amplitudes in any zero mode state, namely all those that cannot be
squeezed from a state satisfying the ${\sf M}+2$-Pauli principle. Note also
that if there exists a zero mode $|\psi\rangle$ at filling factor
$1/({\sf M}+2)$, then the state $|10^{{\sf M}+1}10^{{\sf
M}+1}\dotsc\rangle$ {\em must} have non-zero coefficient in the
expansion of $|\psi\rangle$, and all basis states $|\{n\}\rangle$
appearing in the expansion of $|\psi\rangle$ must  be squeezable from
$|10^{{\sf M}+1}10^{{\sf M}+1}\dotsc\rangle$. This follows since the
latter is the unique basis state satisfying the ${\sf M}+2$ -Pauli
principle at filling factor $\nu\geq 1/({\sf M}+2)$, together with
Corollary 2. It also follows that there  can be at most one zero mode at
filling factor $1/({\sf M}+2)$. For, if there were two, a linear
combination could be formed in which the coefficient of the state
$|10^{{\sf M}+1}10^{{\sf M}+1}\dotsc\rangle$ vanishes. According to the
preceding statement, this is only possible if the  linear combination
vanishes entirely. We thus have the following 

{\bf Corollary 3:}
Let $T_{j1;m}^-$, $m=1,3,\dots,{\sf M}$, be defined as in the Theorem.
If there exists a state $|\psi\rangle$ at filling  factor $1/({\sf
M}+2)$ that is annihilated by all  $T_{j1;m}^-$, $m=1,3,\dots,{\sf M}$,
$j=1/2,\dotsc,L-3/2$, then $|\psi\rangle$ is the unique state with this
property. Furthermore,  the basis states $|\{n\}\rangle$ appearing in
the expansion of $|\psi\rangle$ include the state $|10^{{\sf
M}+1}10^{{\sf M}+1}\dotsc\rangle$, and every such $|\{n\}\rangle$ can be
squeezed from $|10^{{\sf M}+1}10^{{\sf M}+1}\dotsc\rangle$.

For Laughlin states, the latter was observed in Ref.
\onlinecite{haldanerezayi94}. We note once more that the squeezing
principle has been extremely useful in defining a large class of trial
wave functions,\cite{Bernevig1, Bernevig2, ArdonneRegnault11} and that
the associated ``dominance patterns'', or ``root partitions'', from which these states are
squeezed also dominate the thin torus limit \cite{haldanerezayi94,
seidel, berg}, and are furthermore intimately related to ``patterns of
zeros''.\cite{wenwang}
 These patterns contain much useful information,
e.g., concerning quasi-particle statistics. \cite{seidel_statistics}
Many of the states defined through squeezing have, however, not yet been
identified as ground states of a parent Hamiltonian. Our approach is
thus complementary, where we established a squeezing principle for zero
mode states for a class of Hamiltonians 
of the general form Eqs. \eqref{HamiltG}, \eqref{HLLLG}, with, in principle, 
arbitrary coefficients $\eta_k$.
In particular, this is more general than the
usual pseudopotential construction, which is constrained by rotational and translational symmetry.\cite{Haldane}  
Some instances of such more general
Hamiltonians have already surfaced in the recent literature, and have
been shown to exhibit zero modes,\cite{Nakamura} which conform to all
the results of this section.

We emphasize that there is a difference between the well documented
connection between first quantized pseudopotential-type Hamiltonians
and ``clustering properties'' of their analytic ground state wave 
functions,\cite{RRpara, Stone, Bernevig1, Bernevig2, wenwang}
and the approach presented here.
It is well understood how these clustering properties, i.e., certain
analytic properties of first quantized wave functions, are related to squeezing
principles describing their second quantized form.\cite{Bernevig1, Bernevig2, wenwang} 
Here, however, we are {\em not} interested in such clustering properties, which describe certain types of
first quantized wave functions. Indeed, the results given here are {\em not}
limited to cases where first quantized forms of zero modes display such clustering properties.
This is demonstrated, e.g., by the explicit 
examples given in Ref. \onlinecite{Nakamura}, where zero modes
are constructed that satisfy a squeezing principle in accordance
with the results of this section, whilst their first quantized forms do not
display analytic clustering properties.
We note that it is straightforward to
modify our results on a case-by-case basis for situations in which the
independence property is  violated in some form, and to generalize our
results to particles  with spin or internal degrees of freedom.
Likewise, the results and  principles discussed here can be 
generalized  to $n$-body operators, such as the parent Hamiltonians
for states in the Read-Rezayi series.\cite{RRpara}

\subsection{Richardson-Gaudin decomposition of zero energy states}
\label{RG0}

Knowledge of the null space of {\it any}  operator $H_{{\sf G}_{j}}$,
${\sf Ker}(H_{{\sf G}j})$,  helps us find a RG basis to expand
$\ket{\Psi^J_\nu}$; the basis is the set  of zero-energy eigenstates of
$H_{{\sf G}j}$ with fixed $J$.  We next consider  expansion of any
arbitrary zero energy state in terms of this RG basis.  The state with
$J_{\sf m}=N (N -1)/2$ ($\nu=1$),
$\ket{\Psi_1}=\frac{1}{\sqrt{N!}}\prod_{r=0}^{L-1} c^\dagger_r \ket{0}$,
is clearly a {\it unique} eigenstate of $H_{V_1}$ with positive
eigenvalue,  and maximal pairing, meaning that the seniority is zero
(see Table \ref{table:laughlin}). 
\begin{table}[hbt]
\begin{tabular}{ c|c|c|c}
\hline\hline
$\nu$  & $m_{\sf max}=2j_{\sf m}$   &  ${\cal C}(j)$ & $J_{\sf m}$  \\
\hline
$\frac{1}{{\sf q}}$ & $\ {\sf q}(N -1) \ $  & $\ \min
([j+\frac{1}{2}],[L-\frac{1}{2}-j]) \ $ & $\frac{{\sf q}N (N -1)}{2}$ \\
\end{tabular}
\caption{Quantities involved in the description of incompressible Laughlin 
states (see text).}
\label{table:laughlin}
\end{table}

The general second-quantized form of a Laughlin state  with filling
fraction $\nu=1/{\sf q}$ is
\begin{eqnarray}
\ket{\Psi_\nu}=\sum_{M=0}^{N /2} \sum_{\nu(j)}\sum_{\ell=1}^{\ell_M}
 \alpha_{M \nu(j)}^{(\ell)}  \ket{\Phi_{M \nu(j)}^{(\ell)}} ,
\end{eqnarray}
where without loss of generality we assume the total number of electrons
$N $ to be  even, $j_{\sf m}={\sf q} (N -1)/2=(L-1)/2$, and every state
in the sum $\ket{\Phi_{M \nu(j)}^{(\ell)}}$ is  of the form of Eq.
\eqref{statepaired} with total angular momentum $J_{\sf m}=j_{\sf m} N $
and thus  the same filling fraction. The $\nu(j)$ sum is over unpaired
states of a given  seniority $N_{\nu(j)} = N  - 2 M$.  The extra index
$\ell$ labels a particular solution  of the RG equations, Eq.
\eqref{RGeqs}, which for a fixed $M$ has a total of  $\ell_M=
\binom{{\cal C}(j)}{M}-\binom{{\cal C}(j)}{M-1}$  solutions (see  Table
\ref{table:dimensions}). The coefficients $ \alpha_{M \nu(j)}^{(\ell)} $
can be determined by  solving the set of  equations
\begin{eqnarray}
\langle \Phi_{M \nu(j)}^{(\ell)} | \widehat{H}_{{\sf QH}}
\ket{\Psi_\nu} = 0 .
\end{eqnarray}
The RG expansion is similar in spirit, but different from the expansion 
in terms of squeezed Slater determinants,\cite{Francesco,
Bernevig1,Bernevig2}  and can be applied also in general situations
where one wishes to test for the  existence of zero modes in the absence
of a known root partition.

To make our discussion lucid, we now turn to a simple explicit
illustrative  example,  that of $N=2$ particles with ${\sf q}=3$. The QH
Hamiltonian, in this case, is given by
\begin{eqnarray}
H_{V_1}= g \sum_{j=\frac{1}{2},1,\frac{3}{2},2,\frac{5}{2}}T^{+}_{j
1}T^{-}_{j 1} ,
\end{eqnarray}
with $L=4$ orbitals, and $J_{\sf m}=3$. The Hilbert space is spanned  by
${\cal P}(2,4)=2$ eigenstates. The ground state corresponds to the Laughlin
state ($M=1, \nu(\frac{3}{2})=\{0,0\}$) given (in an un-normalized form)
by the eigenvector
\begin{eqnarray}
\ket{\Psi_{\frac{1}{3}}}
=\frac{\eta_{\frac{1}{2}}}{\eta_{\frac{1}{2}}^2-E_1}
c^\dagger_2c^\dagger_1 \ket{0} +
\frac{\eta_{\frac{3}{2}}}{\eta_{\frac{3}{2}}^2-E_1}
c^\dagger_3c^\dagger_0  \ket{0},
\end{eqnarray}
with ${\cal E}=0$  eigenvalue, and  where $E_1$ satisfies the RG
equation
\begin{eqnarray}
\frac{1}{\eta^2_{\frac{1}{2}}-E_1} + \frac{1}{\eta^2_{\frac{3}{2}}-E_1}
 +\frac{2}{E_1} =0,
\end{eqnarray}
whose unique ($\ell_1=1$) solution is $E_1=2
\eta^2_{\frac{1}{2}}\eta^2_{\frac{3}{2}}/(\eta^2_{\frac{1}{2}}+
\eta^2_{\frac{3}{2}})>0$. Associated with the positive energy, ${\cal
E}=g (\eta^2_{\frac{1}{2}}+\eta^2_{\frac{3}{2}})= g$  (See Eq.
(\ref{sumpi}) more generally), there is an eigenvector 
\begin{eqnarray}
T^+_{\frac{3}{2}1} \ket{0}
=\eta_{\frac{1}{2}} c^\dagger_2c^\dagger_1 \ket{0} +
\eta_{\frac{3}{2}} c^\dagger_3c^\dagger_0  \ket{0},
\end{eqnarray}
orthogonal to $\ket{\Psi_{\frac{1}{3}}}$, and corresponding to 
$E_1\rightarrow \infty$.
This particular example constitutes an equivalent of the
{\it unbound}  Cooper pair problem for $\nu=1/3$. We would like to point
out that  the above two eigenvectors are also zero seniority 
eigenstates  of the RG
Hamiltonian $H_{{\sf G}j_{\sf m}}$ with $j_{\sf m}=3/2$.

\subsection{Slater Decomposition of Laughlin states and the role of pairing}
\label{slatersec}

Our generalized RG approach emphasizes the role of pairing. Thus far, we
focused attention on 
the use of the Gaudin algebra, which directly
captures the underlying algebraic structure of the problem, 
and worked as much as possible in a second quantized language. This section is an
exception  where we deliberately make contact with the more traditional first
quantized language. It is illuminating to examine tendencies towards
pairing within a far more standard conduit: the Slater decomposition of
the Laughlin states. \cite{Dunne,Francesco,Bernevig1,Bernevig2} 

We have shown above that finding the zero modes of
pseudopotentials  can be viewed as a frustrated pairing problem, where
the  Hamiltonian is the sum of (mostly) non-commuting pairing terms, 
each of which couples to pairs at different total angular momenta. 
Still, as we will elaborate below, pairing with angular momentum 
$2j_{\sf m}=L-1$ plays a special role, in the sense that both before 
and after normalization, the states of highest amplitude in this
decomposition  are fully paired (or zero seniority) with respect to that
value.

In its Slater decomposition, a $\nu = 1/{\sf q}$ Laughlin state for $N $
spin-polarized electrons, omitting Gaussian prefactors, may be expressed
in a first quantized language as  
\cite{Dunne,Francesco,haldanerezayi94}
\begin{eqnarray}
\label{slatdet}
\hspace*{-0.5cm}
\Psi_\nu(\{z_i\}) =\prod_{1 \leq i, j \leq N } \!\!\!\!
(z_i-z_j)^{\sf q}=\sum_{\{m_i\}}  C^{N }_{\{m_i\}} \prod_{i=1}^{N }
z_i^{m_i} ,
\end{eqnarray}
where $\{m_i\}=\{m_1,m_2,\cdots, m_{N }\}$, $\sum_{i=1}^{N }m_i=J_{\sf
m}$,  and the  coefficients in the expansion, $C^{N }_{\{m_i\}}$, are
integers.  The total number of Slater  determinants needed in the
expansion of $\Psi_\nu(\{z_i\})$ is smaller than $N_{\{m_i\}}$. Direct
relations exist  between the Slater matrix decomposition of the Laughlin
states and Young tableaux and further related aspects such as  the
geometry of high dimensional polytopes. It is noteworthy that not all of
the partitions actually appear in the expansion \eqref{slatdet}. In
particular, only  \cite{Francesco,haldanerezayi94,Bernevig1,Bernevig2}
those Slater determinants appear that can be obtained from the ``root
partition'' by ``inward squeezing''. We have explained and
generalized some of  these notions from a Hamiltonian point of view in
Section \ref{inward}  in the context of second quantization.

For self-completeness, we now briefly review these terms and associated
rudiments.  The Slater determinant basis decomposition is identical to
that carried out in other works using squeezed states represented by one
dimensional strings of ones and zeros to denote viable
states.\cite{Bernevig1,Bernevig2} Any set of integers $\{ m_{i} \}$ in
Eq. (\ref{slatdet}) corresponding to a particular product term
$\prod_{i=1}^{N } z_i^{m_i}$ can be written  as a binary string of ones
and zeros where the ones in the string  appear at the locations $\{m_{i}
\}$. To make this clear,  consider the decomposition of simple
two-particle Laughlin states,
\begin{eqnarray}
(z_{1} - z_{2})^{3} &=& 
   \left| \begin{array}{cc}
 z_{1}^{3} & z_{2}^{3}  \\
1& 1   \end{array} \right| 
-3   \left| \begin{array}{cc}
 z_{1}^{2} & z_{2}^{2}  \\
z_{1}& z_{2}   \end{array} \right|, \nonumber
\\ (z_{1} - z_{2})^{5} &=&  
 \left| \begin{array}{cc}
 z_{1}^{5} & z_{2}^{5}  \\
1& 1   \end{array} \right| 
-5 \left| \begin{array}{cc}
 z_{1}^{4} & z_{2}^{4}  \\
z_{1}& z_{2}   \end{array} \right| 
+ 10 \left| \begin{array}{cc}
 z_{1}^{3} & z_{2}^{3}  \\
z_{1}^{2}& z_{2}^{2}   \end{array} \right|, \nonumber
\\ \! \! \! \! \!  &\vdots& . \nonumber
\end{eqnarray}
Any Slater determinant which appears in such Laughlin state 
decompositions can be expressed as a binary string following a 
well-known schematic which we now review. As an example, consider the 
determinants associated with the $\nu=1/5$ state. The Slater
determinant 
\[ \left| \begin{array}{cc}
 z_{1}^{5} & z_{2}^{5}  \\
1& 1   \end{array} \right|, \] 
i.e.,  the determinant of $z_{1,2}$ raised to the zero and fifth powers
can be denoted by the string $| 1 0000 1 0000 \cdots \rangle$. That is,
in this schematic, there are ones at the zeroth and  fifth entries of
the string (assuming that the leftmost entry of the  string corresponds
to the  ``zeroth" entry).  Similarly, 
\[
\left| \begin{array}{cc}
 z_{1}^{4} & z_{2}^{4}  \\
z_{1}& z_{2}   \end{array} \right|, 
 \left| \begin{array}{cc}
 z_{1}^{3} & z_{2}^{3}  \\
z_{1}^{2}& z_{2}^{2}   \end{array} \right| \]
can be symbolically denoted as  $| 0100100 \cdots \rangle$ and $|001100
\cdots \rangle$. The states $|  0 1001 000 \cdots \rangle, |001100
\cdots \rangle$ can be obtained by ``inward squeezing'' of the ``root
partition'' $| 1 0000 1 0000 \cdots \rangle$  of the $\nu = 1/5$
state. By ``inward squeezing'', we allude to the displacement  of the
pair of ones in $| 1 0000 1 0000 \cdots \rangle$ such that their total
angular momentum (the sum of the powers  of $z_{i}$) is preserved. Now,
here is the important point about ``inward squeezing''. All admissible
Slater determinant states in the decomposition of the Laughlin state are
either a root state (such as $|1 0000 1 0000 \cdots \rangle$) or states
that can be derived by inward squeezing operations from that state. The
root states adhere to a generalized Pauli principle: The ones in the
binary string must be separated by, at least, $({\sf q}-1)$ zeros. 
Thus, the densest root state corresponds to a string such as $|1 00 1 00
\cdots \rangle$ for $\nu =1/3$ or to $|1 00 00 1 00 00 \cdots \rangle$
for $\nu = 1/5$, etc. These configurations have intimate connections to
states appearing in the thin torus limit.  \cite{haldanerezayi94,
seidel, berg} Any state that cannot be derived from the root state
(i.e., a non-admissible state) has a vanishing amplitude in the
decomposition of Eq. (\ref{slatdet}). 

In examining the Slater decomposition we found that for general filling
fractions $\nu$, there may exist a threshold value for the number of
particles $N_{0}(\nu)$ such that when $N \ge N_{0}(\nu)$ there are
admissible states that have a vanishing amplitude. For instance, when
$\nu =1/3$, there exist 
\begin{eqnarray}
N \ge N_{0}(\nu =1/3) = 8
\end{eqnarray}
particle Slater determinant states which albeit being ``admissible'',
from the standpoint of inward squeezing, have a vanishing amplitude in
the  decomposition of the Laughlin states.  The Slater determinants with
non-zero coefficients are in general a true subset of the one obtained
by inward squeezing. \cite{haldanerezayi94} ``Admissible partitions''
\cite{Francesco} were earlier conjectured to all have corresponding
non-vanishing Slater determinant amplitudes; we now see that this
conjecture is incorrect.   We remark that although defined seemingly
differently through Young  tableaux considerations, the admissible
partitions of Ref. \onlinecite{Francesco}  are in fact identical to
those defined via inward squeezing.\cite{haldanerezayi94}  We provide
the simple proof in Appendix \ref{appB}.

References [\onlinecite{Dunne}] and [\onlinecite{Francesco}] provided
explicit forms for the coefficients $C^{N}_{\{m_i\}}$ [see, e.g., Eqs.
(4.11) and (4.22) or Eq. (4.40) of Ref. [\onlinecite{Francesco}]]. 
These may be obtained in a variety of inter-related ways -- all of which
lead to the earlier noted result concerning the dominance of the fully
paired states (in line with the main thesis of our work). We comment on
one recursion relation which enables a computation of the amplitudes
$C^{N }_{\{m_i\}}$. Such a relation may be arrived at by noting that
$\Psi_\nu(\{z_j\}) = \prod_{i=1}^{N -1} (z_{i} - z_{N })^{\sf q}
\Psi_\nu(\{z_j\}_{j=1}^{N -1})$ and expressing the $N $ and $(N -1)$
particle wave functions on both sides of this relation via the Slater
determinant decomposition of Eq. (\ref{slatdet}). This leads to a
recurrence relation between coefficients  with different number of
particles ($m_i^-=m_i-l_i \geq 0$)
\begin{eqnarray}
\hspace*{-0.5cm}
C^{N }_{\{m_i\}} =  (-1)^{m_{N }} \sum_{\{l_i\}} \binom{{\sf q}}{m^-_1}
\cdots  \binom{{\sf q}}{m^-_{N -1}} \ C^{N -1}_{\{l_i\}},
\end{eqnarray}
where $\sum_{i=1}^{N -1}l_i=J_{\sf m}-m_{\sf max}$, and $C^1_0=1$.

When the coefficients $C^{N }_{\{m_i\}}$ are computed it is found that
pairing tendencies prevail. We discuss these explicitly for both (i) the
un-normalized Slater determinants as well as (ii) the decomposition of
the Laughlin state into normalized Slater determinants.

{\bf{(i)}} Un-normalized Slater determinant wave functions. The  largest
coefficient ${C}^{N }_{\{m_i\}}$ is associated with the following 
integer partition
\begin{eqnarray}
\{ m_1&=&\frac{({\sf q}+1)(N -1)}{2}, \,  m_2=m_1-1, \, m_3=m_1-2 , \nonumber
\\
&&, \cdots , \, m_{N }=m_1-(N -1) \} \equiv \{m_{\sf bunch}\},
\end{eqnarray}
which for $N =2M$ represents a state that belongs to the {\it fully
paired} subspace (i.e., that of vanishing seniority)
\begin{eqnarray}
\hspace*{-0.5cm}
m_1+m_{N }=m_2+m_{N -1}=m_3+m_{N -2} \nonumber
\\ =\cdots=2j_{\sf m}.
\end{eqnarray}
Such states were termed ``maximally bunched''  \cite{Dunne} or ``most
compact'' \cite{Francesco} states in earlier works.  The norms of the
coefficients associated with this partition are, for general $m$,
\cite{Francesco} given by
\begin{eqnarray}
|{C}^{N }_{\{m_{\sf bunch}\}}| = \frac{(({\sf q}+1)N  /2)!}{(
(({\sf q}+1)N /2)!)^{N } N !}.
\end{eqnarray}
Several additional properties of ${C}^{N }_{\{m_i\}}$  are noteworthy.
These include a symmetry,
\begin{eqnarray}
{C}^{N }_{\{m_i\}} = {C}^{N }_{\{m_{\sf max}-m_{i}\}},
\end{eqnarray}
where $m_{\sf max} = {\sf q}(N-1)=L-1$ as well as the value of the
coefficients for equally distributed  ``most extended'' 
\cite{Francesco} states (forming densest root states discussed above) 
with natural ``Tao-Thouless'' type renditions
\cite{zhou12,TT,seidel,berg,haldanerezayi94}
\begin{eqnarray}
 {C}^{N }_{m_{\sf max}, m_{\sf max}-{\sf q}, m_{\sf max}-2{\sf q},
 \cdots, 0} =1.
\end{eqnarray}

{\bf{(ii)}} Normalized Slater determinant states. What is of greatest
pertinence are not the bare coefficients ${C}^{N }_{\{m_i\}}$ of Eq.
(\ref{slatdet}) but rather the coefficients that appear with normalized
electronic wave functions. It  is thus appropriate to study the
asymptotic (in $N $) behavior of the coefficients
\begin{eqnarray}
\tilde{C}^{N }_{\{m_i\}} =  \sqrt{m_1! \, m_2! \cdots \, m_{N }!} \
C^{N }_{\{m_i\}}.
\end{eqnarray}
Following this normalization, the states of the highest weight are those
associated with nearly uniformly spaced root states (and their
Tao-Thouless renditions) \cite{zhou12,TT,seidel,berg} followed by an
inward squeezing of the states of highest and lowest angular momenta.
That is,
\begin{eqnarray}
\tilde{C}^N_{m_{\sf max}-1, {\sf q}(N -2), {\sf q}(N -3), \cdots, {\sf q}, 1}
\end{eqnarray}
is the largest amongst all coefficients in the expansion of the Laughlin
wave function in terms of normalized wave functions. \cite{Dunne} This
(as well as, the lower amplitude, uniformly spaced) state has a weight
that increases  exponentially relative to that of the  maximally bunched
state. \cite{Dunne,Francesco} In accord with the main theme of our work,
it is important to note that {\it this largest amplitude state}, i.e., the
state $|m_{\sf max}-1, {\sf q}(N -2), {\sf q}(N -3), \cdots, {\sf q}, 1
\rangle$, {\it  is a state of zero seniority, i.e., a fully paired
state}.

\section{Second-quantized form of quasi-hole generators}
\label{2ndhole}

A remarkable feature of QH Hamiltonians is the fact that in addition to
an incompressible frustration free ground state, they posses many other
zero energy (and hence likewise frustration free) states describing
quasi-hole excitations. The incompressible state is characterized as
having the smallest (angular) momentum, for given particle number $N$,
among the zero modes of the Hamiltonian. The number of quasi-hole states
grows exponentially in the difference between $L$ and $N/\nu$, and
counting formulas have been derived for various Hamiltonians  and
geometries.\cite{RR,ardonne,read06, alexkun} While these properties are
traditionally discussed in first quantized language,  our goal here is
to understand as many of these properties as possible in terms of the
algebraic structure emanating from the second quantized versions of
these Hamiltonians, beginning with the operators defined in Eq.
\eqref{Tj}. 

In this section, we take as given the existence of the incompressible
Laughlin states $|\Psi_\nu\rangle$ at filling factor $\nu=1/{\sf q}=1/({\sf M}+2)$ and (angular)
momentum $J_{\sf m}=N(N-1)/(2\nu)$, which are frustration free ground
states of their respective parent Hamiltonians as discussed in Section
\ref{nullspace} for $\nu=1/3$.
We will show how further zero modes associated with quasi-hole states
are  then generated in second quantization. 
That is, focusing at first
on $\nu=1/3$, we will show how the property \eqref{zeromode} for
$|\Psi_\nu^{J_{\sf m}}\rangle$, to wit,
\begin{eqnarray}
 T^-_{j1;m=1}|\Psi_\nu^{J_{\sf m}}\rangle=0\,,
\end{eqnarray}
leads to the existence of other states satisfying the same condition at
smaller filling fraction.

Our strategy builds on the knowledge that, in first quantization,
general quasi-hole states are generated by multiplying the Laughlin
state with an arbitrary symmetric polynomial. \cite{Haldane,Trugman} It
is thus natural to seek second quantized operators whose action
represents the multiplication of the wave function by a member of a
generating system of the symmetric polynomials. The generating system
that is usually given preference in the literature in related contexts
is that of elementary symmetric polynomials
\begin{eqnarray}
s_{\sf t}=\sum_{{\sf i}_1<{\sf i}_2<\cdots<{\sf i}_{\sf t}}z_{{\sf
i}_1}z_{{\sf i}_2} \cdots z_{{\sf i}_{\sf t}} ,  \ \ {\sf
t}=1,2,\cdots,N .
\end{eqnarray}
These, however, are not ideal for the task at hand, as multiplication
with such polynomials is in general not described by a one-body
operator. Instead, we will work with {\it power-sum symmetric
polynomials} of the form 
\begin{eqnarray}\label{p_d}
p_d=\sum_{{\sf i}=1}^N z_{\sf i}^d , \ \ d=1,2,\cdots,N ,
\end{eqnarray}
that are likewise a generating system of all symmetric polynomials in
$N$ variables. Since the right hand side of Eq. \eqref{p_d} is the sum of terms
each of which depends only on one variable each, the multiplication by $p_d$
corresponds to a one-body operator ${\cal O}_d$. It will be sufficient
to express these operators using the normalization conventions of the
$\kappa=0$ cylinder. For other geometries, the proper expressions for
the ${\cal O}_d$ can be obtained from the ones given below by means of
the similarity transformations defined in Section \ref{equivalence}. We have
\begin{eqnarray}\label{Od}
 {\cal O}_d= \sum_{r \geq 0} c^\dagger_{r+d}c^{\;}_r\quad(d>0).
\end{eqnarray}
In lieu of a (simple) proof that this operator facilitates
multiplication of a state with the polynomial $p_d$, we will prove {\em
directly} from the algebra of the operators $t^-_{j;1}$ (appropriate for
the $\kappa=0$ cylinder)
that for any state satisfying $t^-_{j;1}|\psi\rangle=0$ for all $j$,
${\cal O}_d|\psi\rangle$ will be a new state having the same property.
To this end, we note the commutator 
\begin{eqnarray}\label{comm}
   [t^-_{j;1}, {\cal O}_d] = 2 t^-_{j-d/2;1}\,,
\end{eqnarray}
where the right hand side annihilates $|\psi\rangle$ by assumption, and
therefore $t^-_{j;1}$ indeed also annihilates ${\cal O}_d |\psi\rangle$.

In Eq. \eqref{comm}, we have used the convention $t^-_{j;1}\equiv 0$ for
$j<0$,  and have refrained from introducing an upper cutoff $L$ on
orbital indices, thus working with a half-infinite cylinder. It is clear
that if $n_{\sf max}(\psi)$ is the largest occupied orbital in
$|\psi\rangle$ (e.g., $n_{\sf max}(\Psi _\nu^{J_{\sf m}})= (N-1)/\nu$ for
$|\Psi_\nu^{J_{\sf m}}\rangle$), and $n_{\sf max}(\psi)+d<L$, then the action of
${\cal O}_d$ on $|\psi\rangle$  is completely independent of the
presence or absence of such a cutoff, as is the zero mode property of 
${\cal O}_d |\psi\rangle$. Moreover, under the same circumstances the
state ${\cal O}_d |\psi\rangle$ cannot vanish.  This is best seen by
noting that the $n_{\sf max}({\cal O}_d |\psi\rangle)=n_{\sf
max}(\psi)+d$,  and the basis states that are responsible for this
property have coefficients in the expansion of ${\cal O}_d |\psi\rangle$
that cannot vanish (as they are identical to corresponding  non-zero
coefficients in the expansion of  $|\psi\rangle$). We note that $n_{\sf
max}(\psi)$ can be naturally read-off from the thin cylinder limit
\cite{haldanerezayi94, seidel, berg} or dominance
pattern\cite{Bernevig1,Bernevig2} of the state.  We believe that  by
generalizing the result of Corollary 3 of Section \ref{inward} by
systematic use of the operators ${\cal O}_d$, one could recover the
one-to-one correspondence between dominance patterns and zero modes. In
this way the familiar counting of linearly independent zero
modes\cite{RR} for Laughlin states could be reproduced in principle
without reference to polynomials, or assumptions about adiabatic 
continuity in the thin cylinder limit (see Ref. \onlinecite{alexkun} for
an application of the latter method to the derivation of counting
formulas).  However, we will not pursue this  route here further.

It is not difficult to generalize the above considerations to the zero
modes of general pseudopotential Hamiltonians
\begin{eqnarray}
 \label{pseudo}
 \widehat{H}_{\sf QH}=\sum_{0\leq m\leq {\sf M}} g_m H_{V_m}\;,
\end{eqnarray}
where again $m$, ${\sf M}$ are restricted to be even/odd for
bosons/fermions (which we leave understood from now on), and the
coefficients $g_m$ are positive. It is well known that the unique
incompressible (smallest $n_{\sf max}$) zero mode of Eq. \eqref{pseudo}
is the Laughlin state $|\Psi_\nu^{J_{\sf m}}\rangle$ with $\nu=1/({\sf
M}+2)$.\cite{Haldane}  It follows from the discussion at the end of
Section \ref{equivalence} that for the $\kappa=0$ cylinder, the zero
modes of Eq. \eqref{pseudo} can be also characterized by the constraints
\begin{eqnarray}
    t^-_{j;m} |\psi\rangle =0 \quad \forall j, \ \ 0\leq m\leq {\sf M} ,
\end{eqnarray}
with $t^-_{j;m}= \sum_{k(j)} k^m c^{\;}_{j-k} c^{\;}_{j+k}$ as before,
and similarly for other geometries after the appropriate transformation
(Section \ref{equivalence}). Then, we can show just as before that the
action of ${\cal O}_d$ on $|\psi\rangle$ generates further zero modes.
This follows from the simple observation that $[t^-_{j;m}, {\cal O}_d] =
\sum _{0\leq m' \leq m} b_{m'} t^-_{j-d/2;m'}$, with 
$b_{m'}= 2 {m \choose m'}(d/2)^{m-m'}$ for $m=m' \mod 2$
and $b_{m'}=0$ otherwise,
which generalizes Eq. \eqref{comm}.

We finally remark on the formal equivalence between the operators
\eqref{Od} and the operators that are associated with boson creation
operators in the standard dictionary of the bosonization of a chiral
branch of 1D fermions.\cite{Mattis} In the present case, the ``vacuum'',
$|\Psi_\nu^{J_{\sf m}}\rangle$, is different, but  in the limit of large $N$ one
expects to recover the  commutation relation $[{\cal O}_{-d},{\cal
O}_{d'}]=d \, \delta_{d,d'}$ within the zero-mode subspace, familiar
from the fermion density modes in bosonization. 
(A phenomenological argument for this in a similar setting was given in Ref.
\onlinecite{haldanerezayi94}).
At $d\ll N$, the states
created by the operators  \eqref{Od} out of the Laughlin state are thus
edge modes in the chiral boson edge theory of the Laughlin
state.\cite{wen2}  We emphasize, however, that the results of this
section {\it are not limited} to large $N$ or excitations close to the edge
(i.e., $d\ll N$).

\section{Conclusions}
\label{conclusions}

Our focus has been on establishing a systematic second quantized
framework for QH parent Hamiltonians which, it is our hope, will lead to
a better understanding of the algebraic inner workings of these
Hamiltonians and allow the construction of new exactly solvable models
that may be useful in the context of QH physics as well as other
systems. The ultimate goal  is to deeply understand the nature of the
intrinsic topological  quantum order defining QH fluids.  Our
``bottom-up'' approach is diametrically opposite to the traditional
route of working back from the ground states of the QH system to parent
Hamiltonians, where the ground states are obtained either from
analytic/clustering requirements \cite{laughlin, RRpara, Bernevig1, Bernevig2,
wenwang} or from the construction of appropriate conformal field
theories.\cite{Moore}  The latter has been extremely successful in
particular in the construction of non-Abelian topological phases  and
has helped fueling a flurry of activity in topological quantum
computing.\cite{Pachos} Although our study has largely focused on
Abelian Laughlin states and two-body interactions, the ideas that we
introduce may be extended to more general exotic states under current
investigation, and to more complicated $n$-body interactions.

We conclude with a brief synopsis of our second quantized approach and
some of our key results.   Central findings reported in this work
include the following:

{\bf (1)} We established a relation between (i) a broad class of
rotationally symmetric two-body interactions within the LLL and (ii)
integrable hyperbolic RG type Hamiltonians that arise in
$(p_{x}+ip_{y})$ superconductivity.  Specifically, we illustrated that
Haldane pseudopotentials (and their sums) can be expressed as a sum of 
repulsive, in general non-commuting, $(p_{x}+ip_{y})$-type pairing
Hamiltonians.  That is, the QH system can be viewed as such a composite,
or {\it soup},  of strongly-coupled  pairing systems.  

{\bf (2)} We derived and exactly-solved the RG type Hamiltonian 
relevant for QH physics, which we call QH-RG, and  determined 
the complete eigenspectrum and, in particular, its null
space by Bethe Ansatz.

{\bf{(3)}} Building on the frustration free character of the QH
Hamiltonian, we discussed the  ground state of the full QH problem and
the use of the new  RG basis which highlights pairing.

{\bf{(4)}} We studied the size of the Hilbert space associated with the
RG basis and  related this problem to that of trivially constrained
non-interacting fermions. 

{\bf (5)} We proved separability of arbitrary-order Haldane
pseudopotentials, 
and  provided explicit expressions for their second quantized forms in all standard geometries.

{\bf (6)} By explicit construction, we showed  how to exploit the
topological equivalence between different geometries (disk, cylinder,
and sphere) sharing the same topological genus number in the second
quantized formalism through similarity transformations.

{\bf{(7)}} We established a ``squeezing principle'', in second quantized
language, that applies to the zero modes of a general class of
Hamiltonians, which includes but is not limited to Haldane
pseudopotentials. We also showed how one may establish (bounds on)
Òincompressible filling factorsÓ for  those Hamiltonians, thus
illuminating why certain filling factors are special for certain 
classes of Hamiltonians.

{\bf (8)} Building on the properties of  ``bosonic'' symmetric
polynomials, our second quantized formulation enables an explicit form
for quasi-hole generators. The generators that we find inherently relate
to bosonic {\it chiral boundary edge modes} and further make aspects of
dimensional reduction in the QH systems precise.

{\bf{(9)}} We established equivalence between the Young tableaux
approach  to determining the non-vanishing amplitudes in a Slater
determinant decomposition of  Laughlin states, and the squeezed state
approach. We also noted that  there exists a minimal number of particles
beyond which  there still remain vanishing amplitudes even after
applying these rules.  Finally, we highlighted the presence of pairing
in those standard  Slater determinant decomposition of Laughlin states.

\vspace{0.5cm}
\centerline{
{\bf Acknowledgments}}

This work has been partially supported by the National Science
Foundation under NSF Grant No. DMR-1206781 (AS), NSF Grant No.
DMR-1106293 (ZN), and by the Spanish MICINN Grant No. FIS2012-34479.  GO
would like to thank the Max-Planck-Institute  in Garching and, in
particular, Ignacio Cirac for the great hospitality during the final 
stages of the work. AS would like to thank R. Thomale for insightful
discussion of the final manuscript.

\appendix

\section{Derivation of the polynomial structure of the $\eta_{k}(j,m)$ of
$V_m$ for disk and sphere geometries}
\label{appA}

A two-particle state for given relative angular momentum $m$ and total
angular momentum $2j$ is given by Eq. \eqref{jmstate1st}, the polynomial
part of which we reproduce here as
\begin{equation}\label{poly}
 {\cal N} (z_1+z_2)^{2j-m}(z_1-z_2)^m={\cal N}\sum_k C_{mjk}
z_1^{j-k}z_2^{j+k},
\end{equation}
where the normalization constant is ${\cal N}= 
\frac{2^{-2j}}{2\pi\sqrt{(2j-m)! m!}}$. In the monomial expansion on the
right hand side of  Eq. \eqref{poly}, the term $z_1^{j-k}z_2^{j+k}\pm
(1\leftrightarrow 2)$ corresponds to the state ${\cal
N}_{jk}c^\dagger_{j-k}c^\dagger_{j+k}|0\rangle$ where ${\cal
N}_{jk}=2\pi 2^{j+1/2}\sqrt{(j-k)!(j+k)!}$. As usual, the special case
$k=0$ for bosons is taken care of properly by writing $k$-sums as in Eq.
\eqref{jmstate}  in the form \eqref{ksum} and need not be considered
separately. We see that, for given $j$,  $\eta_k$ is the coefficient of
$z_1^{j-k}z_2^{j+k}$ in Eq. \eqref{poly}, multiplied by ${\cal N}_{jk}$:
\begin{eqnarray}\label{etadisk2}
  \eta_k&&=C_{mjk}{\cal N}{\cal N}_{jk} = {\cal N} {\cal N}
  _{jk}(-1)^{m+j-k}\nonumber\\
  &&\times  \sum_{\ell=0}^{j-k} (-1)^\ell {2j-m \choose \ell}{m \choose
  j-k-\ell}
\end{eqnarray}
The sum can be formally written in terms of a hypergeometric function,
which gives
\begin{equation}
\begin{split}
 \eta_k&=  {\cal N} {\cal N} _{jk}(-1)^{m+j-k}\\
 &\times {m \choose j-k} {_2} F_1 (-j+k,-2j+m,1-j+k+m,-1)
\end{split}
\end{equation}
which is Eq. \eqref{hyper}.

We now want to derive the useful alternative expression \eqref{etadisk}.
To this end, we need to show that
\begin{eqnarray}\label{Cmjk}
   C_{mjk}={2j \choose j+k} p_{mj}(k)
\end{eqnarray}
with $p_{mj}(k)$ an $m$th order polynomial in $k$ of parity $(-1)^m$.
This in Eq.  \eqref{etadisk2} gives Eq.  \eqref{etadisk}. We prove Eq.
\eqref{Cmjk} via induction in $2j$. For $2j<m$, we set $C_{mjk}=0$ so
there is nothing to prove. We hence start with $2j=m$. In this case, Eq.
\eqref{poly}  immediately implies
\begin{eqnarray}
   C_{m\frac{m}{2}k}= {m\choose \frac{m}{2}+k} (-1)^{\frac{m}{2}+k}\,.
\end{eqnarray}
Note that $k$ assumes the independent values  $0\leq k\leq m/2$ with $k$
integer/half-odd-integer for $m$ even/odd. Of these there are $m/2+1$
for $m$ even and $(m+1)/2$ for $m$ odd. This exactly equals the number
of free parameters in the polynomial $p_{mj}(k)$. We can thus choose
$p_{m\frac{m}{2}}(k)$ such that
$p_{m\frac{m}{2}}(k)=(-1)^{\frac{m}{2}+k}$ for the indicated $k$ values,
and this proves Eq. \eqref{Cmjk} for  $2j=m$.

Specifically, for $m$ even, we define
\begin{subequations}
\begin{eqnarray}
 q_{ml}= \prod_{\substack{0\leq r\leq m/2\\ r\neq l}} (k^2-r^2)\,,
\end{eqnarray}
and
\begin{eqnarray}
 q_{ml}= k\prod_{\substack{1/2\leq r\leq m/2\\ r\neq l}} (k^2-r^2)
\end{eqnarray}
\end{subequations}
for $m$ odd and $0\leq l\leq m/2$,  again with $2l$, $2r$ restricted to
have  the same parity as $m$.

Then 
\begin{equation}\label{pmm2}
 p_{m\frac{m}{2}}(k)= \sum_{\substack{0\leq l\leq m/2\\ l\in {\mathbb
Z}+\frac{1-(-1)^m}{4}}} (-1)^{\frac{m}{2}+l}\,q_{ml}(k)/q_{ml}(l)
\vspace{3mm}
\end{equation}
satisfies the desired properties. In particular, it is of degree $m$ and
no less, since it must have $m$ zeros in the interval $(-m/2,m/2)$ for
continuity reasons.

We now assume that Eq. \eqref{Cmjk} holds for some value of $2j$.
Multiplying Eq. \eqref{poly} by $(z_1+z_2)$, it is elementary to show
that
\begin{equation}
 C_{m\frac{2j+1}{2}k}=C_{mj\frac{k-1}{2}}+C_{mj\frac{k+1}{2}}\,.
\end{equation}
Using Eq. \eqref{Cmjk}, this gives
\begin{equation}
 C_{m\frac{2j+1}{2}k}={2j+1\choose j+\frac 12
+k}p_{m\frac{2j+1}{2}}(k)\,,
\end{equation}
where we have the recursive relation
\begin{equation}
\begin{split}\label{diskrecursion}
 p_{m\frac{2j+1}{2}}(k)
= \frac 12\left[p_{mj}(k-\frac{1}{2})+p_{mj}(k+\frac{1}{2})\right]\\
+\frac{k}{2j+1}\left [p_{mj}(k-\frac{1}{2})-p_{mj}(k+\frac{1}{2})\right]\,.
\end{split}
\end{equation}
It is manifest from the above expression that if $p_{mj}(k)$ is a
polynomial in $k$ of degree $m$ and parity $(-1)^m$, then so is
$p_{m\frac{2j+1}{2}}(k)$. Specifically, the coefficient of $k^m$ in
$p_{m\frac{2j+1}{2}}(k)$ picks up a factor $(2j+1-m)/(2j+1)$ relative to
that in $p_{mj}(k)$, which is always non-zero for $2j\geq m$.

For the sphere, we may proceed in a highly analogous manner, working
instead with the recursion relations of the $3j$-symbols. In this way,
we find a recursion relation for the polynomials $\tilde p_{m,j}(k)$
defined in Eq. \eqref{etasphere} that differs from \Eq{diskrecursion}
only by an overall $j$-dependent factor:
\begin{widetext}
\begin{equation}
\label{sphererecursion}
 \tilde p_{m\frac{2j+1}{2}}(k)
=\frac{2j+1}{\sqrt{(2N_\Phi-m-2j)(2j+1-m)}}\left( \frac 12\left[\tilde
p_{mj}(k-\frac{1}{2})+\tilde p_{mj}(k+\frac{1}{2})\right]
+\frac{k}{2j+1}\left [\tilde p_{mj}(k-\frac{1}{2})-\tilde
p_{mj}(k+\frac{1}{2})\right]\right)\,.
\end{equation}
\end{widetext}

Moreover, the ``initial values'' also differ from Eq. \eqref{pmm2} by
some extra factors:
\begin{equation}
 \tilde p_{m\frac{m}{2}}(k)= \tilde {\cal N} \sum_{\substack{0\leq l\leq
m/2\\ l\in {\mathbb Z}+\frac{1-(-1)^m}{4}}}
\frac{(-1)^{\frac{m}{2}+l}}{{2N_\Phi-m\choose
N_\Phi-\frac{m}{2}+k}}\,q_{ml}(k)/q_{ml}(l),
\end{equation}
where $\tilde {\cal N} =
\sqrt{2\,\frac{2N_\Phi-2m+1}{2N_\Phi-m+1}{{2N_\Phi-2m \choose
N_\Phi-m}{2N_\Phi\choose N_\Phi}}/{{2N_\Phi \choose m}}}$.

Note that the polynomials $p_{mj}(k)$ and $\tilde p_{mj}(k)$ defined
here are each subject to the orthogonality relation \eqref{etaorth}.

\section{Equivalence between admissible Young Tableaux and Squeezing expansions}
\label{appB}

In the following, we establish a simple equivalence  between notions of
``admissible Slater determinants'' that have appeared in the literature.
In Ref. \onlinecite{Francesco}, it was shown that  only Slater
determinants corresponding to ``admissible Young tableaux'' may have
non-zero coefficient in the expansion of the Laughlin state
$|\Psi_{\frac 1 {\sf q}}^{J_{\sf m}}\rangle$, while the same is
known\cite{haldanerezayi94} for the set of Slater determinants
obtained by inward squeezing from the root state $|1000 \cdots 0 1 0
\cdots 0 1 0 \cdots \rangle$, which has $1$'s at positions $m_k={\sf
q}k$, and $0$'s everywhere else. Here we remark that both sets are
identical (and in general, as we pointed out in Sec. \ref{slatersec},
contain the Slater determinants with non-zero coefficient as a true
subset).

In this appendix, we specialize to fermions (${\sf q}$ odd) and denote
Slater determinants by the set of integers $0\leq m_0<\dotsc<m_{N-1}$
denoting the positions of the $1$'s in the occupation number string.
Then, the $m_k$ corresponding to admissible Young tableaux are of the
form\cite{Francesco}
\begin{eqnarray}
\label{condition_Young_1}
m_{k} = {\sf q} k + \Delta_{k},  \nonumber
\\ \Delta_{k} = {\sf n}_{k+1}- {\sf n}_{k},
\end{eqnarray}
where ${\sf n}_{k}$ are non-negative integers that are subject to the
constraints 
\begin{eqnarray}
\label{condition_Young_2}
{\sf n}_{k} \le \frac{1}{2} ({\sf n}_{k+1} + {\sf n}_{k-1}) + {\sf s}, \
\ \  {\sf n}_{0} = {\sf n}_{N} =0,
\end{eqnarray}
and ${\sf q}=2{\sf s}+1$. It is easy to see that the first of conditions
\eqref{condition_Young_2} just ensures the ordering $m_{k+1}>m_k$.

On the other hand, inward squeezed occupation number  patterns can be
characterized by the following two conditions:\cite{Bernevig1}
\begin{equation}\label{condition_squeeze_1}
   \sum_k m_k = \sum_k {\sf q} k=\frac {\sf q} 2 (N-1)N=J_{\sf m}
\end{equation}
and
\begin{equation}\label{condition_squeeze_2}
   \sum_{k=N-\ell}^{N-1} ({\sf q}k-m_k)\geq 0 \text{ for } \ell=1\dotsc
N-1\,.
\end{equation}
One may see that this definition agrees with the more intuitive
definition of inward squeezing in terms of momentum conserving two
particle processes as described in Section \ref{slatersec}. Condition
\eqref{condition_squeeze_1} implies that the state described by $m_k$
has the same (angular momentum) as the root state, and certainly follows
from Eqs. \eqref{condition_Young_1}, \eqref{condition_Young_2}, since
$\sum_k \Delta_k ={\sf n}_{N} - {\sf n}_{0}=0$. Condition
\eqref{condition_squeeze_2} then assures that $m_k$ can be generated by
squeezing processes that are ``inward''. In terms of Eqs.
\eqref{condition_Young_1} and  \eqref{condition_Young_2}, we have $
\sum_{k=N-\ell}^{N-1} ({\sf q}k-m_k)={\sf n}_{N-\ell}\geq 0$, and thus
it follows that every Slater determinant corresponding to an admissible
Young tableau also belongs to the squeezed set. By reversing the logic,
one easily sees that the converse is also true.

\end{document}